\documentclass[a4paper,fleqn,usenatbib]{mn2e}
\usepackage{color}
\usepackage{epsfig}
\usepackage{epstopdf}
\usepackage{amsmath, amssymb}
\usepackage[a4paper,colorlinks=True,citecolor=blue]{hyperref}
\usepackage{wasysym} 
\usepackage{multirow}
\usepackage{lipsum}
\usepackage{ulem}
\usepackage{setspace}
\usepackage{float}
\usepackage[T1]{fontenc}
\usepackage{aecompl}
\usepackage{graphicx,color,bm}
\usepackage{latexsym}
\usepackage{epsfig}
\usepackage{amssymb}
\usepackage{amsmath}
\usepackage{wasysym}
\usepackage{pstricks}
\usepackage{enumitem}
\usepackage{booktabs}
\usepackage{arydshln}

\def\etal{{\frenchspacing\it et al.}}

\def\be{\begin{equation}}
\def\ee{\end{equation}}
\def\ba{\begin{eqnarray}}
\def\ea{\end{eqnarray}}

\newcommand{\ihMpc}{h{\rm\;Mpc^{-1}}}

\def\LaTeX{L\kern-.36em\raise.3ex\hbox{a}\kern-.15em
    T\kern-.1667em\lower.7ex\hbox{E}\kern-.125emX}

\begin{document}

\voffset-1.25cm
\title[Measuring the evolution of  the growth rate ]{The clustering of the SDSS-IV extended Baryon Oscillation Spectroscopic Survey DR14 quasar sample: measuring the evolution of
  the growth rate using redshift space distortions between redshift 0.8 and 2.2}
\author[Ruggeri \etal]{
\parbox{\textwidth}{
%----------------------------------------------
Rossana Ruggeri$^{1}$\thanks{Email: rossana.ruggeri@port.ac.uk},
 Will J. Percival $^{1}$, H\'ector Gil-Mar\'in$^{2,3}$, Florian Beutler$^{1,4}$,  Eva-Maria Mueller$^{1}$, Fangzhou Zhu $^{5}$, Nikhil Padmanabhan$^{5}$, Gong-Bo Zhao $^{6,1}$, Pauline Zarrouk$^{7}$,  
            Ariel G. S\'anchez $^{8}$,
                  Julian Bautista$^{9}$,
         Jonathan Brinkmann$^{10}$,
            Joel R. Brownstein$^{9}$,  
             Falk Baumgarten$^{11}$, 
    Chia-Hsun Chuang$^{12, 11}$,
                   Kyle Dawson$^{9}$,
                  Hee-Jong Seo$^{13}$,
                       Rita Tojeiro$^{14}$,
                     Cheng Zhao$^{15}$ 
}
%
%  et eBOSS collaboration}
\vspace*{15pt} \\
$^{1}$ Institute of Cosmology \& Gravitation, University of Portsmouth, Dennis Sciama Building, Portsmouth, PO1 3FX, UK\\
$  ^2$ Sorbonne Universites, Institut Lagrange de Paris (ILP), 98 bis Boulevard Arago, 75014 Paris, France \\
$  ^3$ Laboratoire de Physique Nucleaire et de Hautes Energies, Universit ´ e Pierre et Marie Curie, 4 Place Jussieu, 75005 Paris, France\\
$  ^4$ Lawrence Berkeley National Lab, 1 Cyclotron Rd, Berkeley CA 94720, USA\\
$  ^5$ Dept. of Physics, Yale University, New Haven, CT 06511\\
$  ^6$ National Astronomy Observatories, Chinese Academy of Science, Beijing, 100012, P.R.China,\\
$^{7}$ IRFU,CEA, Universit\'e Paris-Saclay, F-91191 Gif-sur-Yvette, France \\
$^{8}$   Max-Planck-Institut f$\ddot{u}$r extraterrestische Physik, Postfach 1312, Giessenbachstr., 85741 Garching, Germany \\
$^{9}$  Department of Physics and Astronomy, University of Utah, 115 S. 1400 E., Salt Lake City, UT 84112, USA. \\
$^{10}$ Apache Point Observatory, P.O. Box 59, Sunspot, NM 88349\\
$^{11}$ Leibniz-Institut f$\ddot{u}$r Astrophysik Potsdam (AIP), An der Sternwarte 16, D-14482 Potsdam, Germany.\\
$^{12}$ Kavli Institute for Particle Astrophysics and Cosmology \& Physics Department, Stanford University, Stanford, CA 94305, USA \\
$^{13}$  Department of Physics and Astronomy, Ohio University, Clippinger Labs, Athens, OH 45701\\
$^{14}$ School of Physics and Astronomy, University of St Andrews, North Haugh, St Andrews KY16 9SS, UK. \\
$^{15}$ Tsinghua Center for Astrophysics and Department of Physics, Tsinghua University, Beijing 100084, China.\\
}

\date{\today} 
\pagerange{\pageref{firstpage}--\pageref{lastpage}}

\label{firstpage}

\maketitle

%\clearpage

\begin{abstract} 
  We measure the growth rate and its evolution using the anisotropic
  clustering of the extended Baryon Oscillation Spectroscopic Survey
  (eBOSS) Data Release 14 (DR14) quasar sample, which includes
  $148\,659$ quasars covering the wide redshift range of
  $0.8 < z < 2.2$ and a sky area of $2112.90$ $\rm deg^2$. To optimise
  measurements we deploy a redshift-dependent weighting scheme,
  which allows us to avoid binning, and perform the data analysis
  consistently including the redshift evolution across the sample.  We
  perform the analysis in Fourier space, and use the redshift evolving
  power spectrum multipoles to measure the redshift space distortion
  parameter $f\sigma_8$
% alongside nuisance parameter, 
and parameters
  controlling the anisotropic projection of the cosmological
  perturbations. We measure $f \sigma_8(z=1.52)=0.43 \pm 0.05 $ and
  $df\sigma_8/dz (z=1.52)= - 0.16  \pm 0.08$, consistent with the
  expectation for a $\Lambda$CDM cosmology as constrained by the
  Planck experiment.
\end{abstract}


\section{Introduction}  \label{sec:intro}
\label{sec:intro}

The positions of galaxies signpost peaks in the density field, and
consequently measuring their clustering provides a wealth of cosmological
information. Two components of the clustering are particularly
important: Baryon Acoustic Oscillations (BAO) act as a robust standard
ruler, allowing geometrical measurements from measurements of their
projected sizes, while Redshift-Space Distortions (RSD) change the
clustering amplitude in a way that is anisotropic around the
line-of-sight. The strength of the RSD signal depends on the rate of
structure growth at the redshifts of the galaxies, and therefore
allows tests of General Relativity on extremely large scales. The
combination of these measurements is able to distinguish between
competing models of Dark Energy, the phenomenon driving the
accelerated expansion of the Universe.

Dark Energy starts to dominate the Universe at a redshift
$z\sim0.7$ and, in order to understand the physics behind this in
detail, we desire BAO and RSD measurements covering a wide range of
redshifts. In particular, measurements at redshifts significantly
greater than $0.7$ allow us to measure the amplitude of fluctuations
before Dark Energy dominates, normalising measurements of acceleration
at lower redshifts. The extended Baryon Oscillation Spectroscopic
Survey (eBOSS; \citealt{2016dawson}), part of the SDSS-IV experiment
\citep{2016blanton} was designed with this specific goal in mind
\citep{2016Gongbo}, with the dominant target for observations being
quasars between the redshifts of $0.8<z<2.2$, at a relatively low
density of $82.6\rm/deg^{2}$.

We expect significant evolution in such a sample with redshift: for
example, the bias of these quasars is expected to evolve as
$b(z)\simeq0.28[(1+z)^2-6.6]+2.4$ \citep{2005croom,2017laurent}, thus
ranging from $1.6$ to $3.4$ across the survey. Consequently, when
analysing data we need to be careful to allow for this
evolution, both when optimising any kind of analysis as well as to make
sure measurements are unbiased. The method of ``redshift-weights''
does this by constructing sets of weights applied to all of the data,
before calculating clustering statistics (such as the power spectrum
multipoles). The weights are designed to allow the optimal measurement
of evolving cosmological parameters. The cosmological parameters could
be, for example, the coefficients of a Taylor expansion of the growth
rate with redshift.

\citet{zhu2014}, \citet{2016Mio}, and \citet{2017eva} calculated and
analysed weights optimised to measure the distance-redshift relation
from BAO, the growth rate from RSD, and primordial non-Gaussianity
from the large-scale bias respectively. Recently, these ideas were
applied to mock catalogues for BAO \citep{2016Zhu} and RSD
\citep{2017Mio}, demonstrating their potential. The technique is
now ready to be applied to data, and the characteristics of the eBOSS
quasar sample make it the ideal choice for such analysis. In a companion paper,
\citet{2017Zhu}, a similar technique is applied to measure the BAO,
whereas we instead focus here on RSD measurements. In
\citet{2017Gongbo} and \cite{2017DanDan}, a different methodology is
used to measure the evolving RSD and BAO signals: standard
measurements are made as if for a narrow redshift interval, but
instead for weighted distributions of the quasars. A cosmological
model can be tested by using the supplied sets of weights to determine
the effective RSD and BAO in the model given that kernel, and
comparing to the corresponding measurements.
 
In our paper, we apply the method presented in \citet{2017Mio},
and consider two sets of weights designed to test for deviations from
the $\Lambda$CDM model, by altering $\Omega_m(z)$, or $f
\sigma_8(z)$. The first choice can change both growth and geometry,
unless we explicitly fix one of these, while the second only tests the
cosmological growth rate. We also consider a traditional analysis,
where we only apply weights matching those of \citet{FKP}. This
corresponds to a limit of the redshift-weighting approach as the
redshift-weights tend towards the FKP form in the limit where the
error associated with a cosmological parameter does not vary with
redshift. Our paper is laid out as follows: In
Section~\ref{sec:survey} we briefly review the eBOSS
data. Section~\ref{sec:method} provides an overview of the method,
focussing on the eBOSS specific aspects. The results are presented in
Section~\ref{sec:results}, and discussed in
Section~\ref{sec:discussion}.

\section{The eBOSS DR14 dataset}  \label{sec:survey}

The eBOSS survey \citep{2016dawson, 2016Gongbo} will provide a
redshift survey covering the largest volume to date at a density where
it can provide useful cosmological measurements.  Full survey details
can be found in \cite{2016dawson}: observations will ultimately
include $250,000$ luminous red galaxies (LRGs), $195,000$ emission
line galaxies (ELGs) and over $500,000$ quasars. The main goal is to make
BAO distance measurements at 1--2\% accuracy \citep{2016Gongbo}.
Using the same samples the goal for the RSD analysis is to  constraint $f\sigma_8$ at  $2.5\%$,
$3.3\%$ and $2.8\%$ accuracy for LRGs, ELGs and Clustering Quasars respectively.
For the current analysis we make use of the quasar catalogues from
the eBOSS DR14 \citep{dr14} dataset. The target selection algorithm is
presented in \cite{2016Myers}: quasars were selected from the
combination of SDSS imaging data \citep{dr8}, and that from the WISE
satellite \citep{WISE}. The SDSS imaging data were taken using the
Sloan telescope \citep{Gunn98,Gunn06}, and spectra were taken using
the BOSS spectrographs \citep{2013smee}. Redshifts were measured using
the standard BOSS pipeline \citep{2012bolton}, coupled with various
updates and visual inspection of a subset as outlined in
\citet{dr14}, which describes the DR14Q quasar catalogue.

The quasar sample, covers a wide redshift range, $0.8<z<2.2$ with a
low density, $82.6\rm/deg^{2}$, compared with other targets, and is
designed to ultimately cover a total area of 7500\,$\rm deg^2$. In this
paper we use the intermediate data sample referred to as DR14
\citep{dr14}. This sample contains 98577 quasars covering the wide
redshift range of $0.8 < z < 2.2$ and a sky area of $1001.25$
$\rm deg^2$. Early measurements of the bias of this sample are
presented in \citet{2017laurent}, showing excellent agreement with those
measured from earlier catalogues \citep{2005croom}.  In this work we
make use of the \textit{fiducial} redshift estimates, obtained as a
combination of the three  different estimates ($z_{MgII}$, $z_{PCA}$,
$z_{PL}$), presented in \citet{dr14} and we show the constraints
obtained when measuring the full NGC + SGC samples.  The comparison
between the results from different redshift estimates and the
discussion for the analysis on NGC (SGC) only is presented in \cite{2017Zarrouk}, \cite{2017Hector}
%\ref{Zarrouk17, Hector}.

We apply a number of weights in order to correct for various features
of the data. First, we apply a set of systematics weights designed to correct for
trends observed in the target catalogue, where the density of targets
varies with observational parameters. These weights are presented in
\citet{2017Zarrouk} and our  treatment is consistent with this. %\citet{2017Zarrouk}.
We upweight the nearest neighbour to correct for
close-pairs.
% \textbf{add how many close pairs we have}. 
Redshift failures are corrected by downweighting the random
catalogue used to define the survey mask, as a function of the plate
position: which alters the expected signal-to-noise
\citep{2017Zarrouk}. 
In addition, we apply redshift-dependent weights optimised to measure
the value and derivative of a cosmological parameter (chosen to be
$\Omega_m(z)$ or $f\sigma_8(z)$) beyond a fiducial $\Lambda$CDM model,
around a pivot redshift. The design of these weights considers the
information available and the dependence on the cosmological
parameter of interest. For the eBOSS quasar data, it is not useful to
probe beyond the first derivative of the parameters  around a
pivot redshift because of the limited constraining power of the data set. The derivation of the weights was presented in
\citet{2017Mio}. 
%Note that %
%while the above weights correct for survey incompleteness, this weights attempt to optimise the $S/N$ and to properly
%

In the following sections we briefly review the
key points of the analysis.

\section{Modelling the data}  \label{sec:method}

We contrast three methods:
\begin{enumerate}
\item A traditional analysis, fitting with one set of weights, matching
  those introduced by \cite{FKP}, commonly known as FKP
  weights,  \label{method1}
\item Redshift-weighted, with two sets of weights optimised to measure
  $\Omega_m(z)$; we refer to this method also as $w_{\Omega_m}$. \label{method2}
\item Redshift-weighted, with two sets of weights optimised to measure
  $f\sigma_8(z)$. we refer to this method also as $w_{f\sigma_8}$ \label{method3}
\end{enumerate}
We perform fits either allowing the anisotropic geometrical
projection parameters (also know as the AP parameters \citep{1979AP}, $\alpha_\parallel$ and $\alpha_\perp$ to be
simultaneously fitted, or keeping them fixed at their fiducial value. 

We derive and fit models for all three of these methods using the same
procedure, as described in \cite{2017Mio}. Briefly, we calculate
the TNS model \citep{2010TSN} for each model to be tested at a discrete
series of redshifts and apply the redshift weights to give models of
the redshift-space moments. In order to account for the coupling
between redshift evolution in the cosmological parameters and the
survey geometry on the power spectra moments we discretise the window
convolution, creating sub-windows at redshifts 0.87, 1.01, 1.15, 1.29,
1.43, 1.57, 1.71, 1.85, 1.99, 2.13, following the procedure described
in \cite{2016Mio}.

The TNS model requires us to calculate the non-linear matter power
spectra, $P_{\delta\delta}$, $P_{\delta\theta}$, $P_{\theta\theta}$,
which we do at $1$-loop order in standard perturbation theory (SPT)
using the linear power spectrum input from CAMB \citep{Lewis2002ah}.
% For method~\ref{method1}
%\&~\ref{method2}, we fix $\sigma_8(z)$ at the value of the fiducial
%model when calculating the non-linear contributions.

Quasar bias is modelled including non-local contributions
\citep{2012bias, 2012bald}, with parameters corresponding to the
linear $b$, second order local $b_2$, non local $b_{s2}$, and the
third order non-local $b_{\rm 3nl}$ bias parameters. Given the lack of
sensitivity of the quasar data, we can make  the approximations
$b_{s2}=-4/7(b-1)$ and $b_{\rm 3nl}= 32/315 (b-1)$ following
\citet{2012bald} and \citet{2014saito} respectively.  We assume $b$
linearly evolves with redshift, and that $b_2$ does not vary with
redshift. In fact we know that the bias evolves strongly with redshift
\citep{2005croom,2017laurent} but, given that we wish to constrain
cosmological evolution across the sample, this should be
simultaneously fitted with the cosmological measurements to avoid
double-counting information. We perform a linear fit to match the
linear cosmological measurements as a Taylor expansion with respect
the value of $b\sigma_8$ at the pivot redshift,
\begin{equation}  \label{bz}
 b\sigma_8(z) = b\sigma_8(z_p) + \partial b\sigma_8 /\partial z|_{z_p} (z -z_p).
\end{equation}
With $b\sigma_8(z_p)$ and $\partial b\sigma_8/\partial z|_{z_p}$ free parameters.
We also fit for a constant shotnoise term $S$.

The \textit{traditional analysis}, method~\ref{method1}, makes
measurements at a single effective epoch ($z_{piv} = 1.52$), using
only FKP weights, so we have a single weighted monopole moment, and a
single weighted quadrupole moment to be fitted with five free parameters
in total: $f\sigma_8$, $b\sigma_8$, $\sigma_{\rm Fog}$, $b_2\sigma_8$,
$S$. When allowing the background geometry to vary, this parameter set
is extended to seven, $f\sigma_8$, $b\sigma_8$, $\sigma_{\rm Fog}$,
$b_2\sigma_8$, $S$, $\alpha_\parallel$, $\alpha_\perp$, including the
projection parameters. To validate this model we fitted to a single
snapshot drawn from the Outerim simulation \citep{2016outrim}, with results presented in \cite{2017Zarrouk}. Good
agreement was recovered. We compare our traditional measurement with
other results obtained from similar analyses in  \cite{2017Zarrouk, 2017Hou, 2017Hector, 2017Gongbo}
%\textbf{(REF otheranalyses)}.

Method~\ref{method2} explores deviations from $\Lambda$CDM
through the evolution of $\Omega_m$ in redshift. To do so we model
$\Omega_m(z)$ as a Taylor expansion about the fiducial model
$\Omega_{m, \rm fid}$,
\begin{equation}
\Omega_m(z) = \Omega_{m, \rm fid} q_0 [1 + q_1 y (z)] 
\label{omz}
\end{equation} 
with $y(z)=\Omega_{m, \rm fid}(z)/\Omega_{m, \rm fid}(z_{piv})$; $q_0$
and $q_1$ are free parameters giving the overall normalisation and
first derivative of $\Omega_m(z)$ at the pivot redshift. In this work,
we use a pivot redshift $z_{\rm piv} = 1.52$, matching the effective
redshift of the quasar sample.  To test the robustness of the analysis we
perform the same analysis selecting $z_p = 1.1; \; z_p = 1.7$
confirming that there is no dependence on the pivot redshift selected;
For this method we have two sets of weights for the monopole and two
sets of the quadrupole, so we simultaneously fit to four moments in
total.

This parameterisation provides a common framework to test for
deviations from the fiducial cosmology both in terms of geometry
(distance-redshift relation) and growth rate ($f\sigma_8$), by writing
these quantities as a function of $\Omega_m(q_0, q1)$: we assume that,
for small deviations, we can still assume the standard equations
linking the Hubble parameter and the Angular Diameter distance to
$\Omega_m(z)$, as in the $\Lambda$CDM model. This is discussed further
in \citet{2016Mio}.  Once we have measured $q_0$, and $q_1$, we can
project them back to $\alpha_\parallel$, $\alpha_\perp$ and
$f\sigma_8$ at any epoch. The physical limit that $\Omega_m$ cannot be
negative at any epoch  places a physical motivated prior on
$\alpha_\parallel$, $\alpha_\perp$ and $f\sigma_8$; the impact of such
priors is discussed in detail in Section \ref{ompriors}

The third parametrisation, method~\ref{method3} explores the evolution
of $f\sigma_8$; it represents a more direct way to measure deviations
in structure growth, where the latter are artificially kept separate
from the geometrical evolution.  Here we directly Taylor expand
$f\sigma_8(z)$:
\begin{equation}
  [f\sigma_8](z) = [f\sigma_8]_{\rm fid}(z) p_0 [1 + p_1 x(z)],
\label{fsz}
\end{equation}
where $x=[f\sigma_8]_{\rm fid}(z)/[f\sigma_8]_{\rm fid}(z_{piv})$, and
$p_0$ and $p_1$ are free parameters giving the overall normalisation
and first derivative of $f\sigma_8(z)$ at the pivot redshift. This
model allows a wider range of deviations from the $\Lambda$CDM
scenario, as it does not assume any particular form or relation for
$f$ and $\sigma_8$. For this method we have two sets of weights for
the monopole and two sets of the quadrupole, so we simultaneously fit
to four moments in total.

In \citet{2017Mio}, we compare the \textit{traditional analysis}
to the measurement from the redshift weights techniques projected at
the pivot redshift using mock catalogues, confirming that the redshift
weights analysis give unbiased constraints.  Weights optimised to look
for deviations from $\Lambda$CDM using changes in either $\Omega_m$ or
$f\sigma_8$, provide complementary measurements given the different
deviations, and dependencies on observations. Both can be used to
measure $f\sigma_8$ at any particular redshift, and be compared to the
more traditional way of looking for deviations.

\section{Fitting models to the data}

We now fit to the quasar data with each of the three models, traditional,
$\Omega_m$, $f\sigma_8$, described in Section \ref{sec:method}. We fit
to the NGC and SGC data independently, assuming they are uncorrelated,
a reasonable assumption given their physical separation, and then
combine the likelihoods to give our result from the full NGC + SGC
sample. The results presented in the following sections have been
obtained by simultaneously fitting the full set of parameters using a
MCMC approach, and then marginalising over the parameters not plotted
or measured, including the nuisance parameters $S$ and $\sigma_{\rm Fog}$,
common to all methods.

We measure the weighted moments of the power spectrum, using the
method described in \cite{bianchi2015}, with different sets of
weights. We select 30 $k$-bins, $0.001<k<0.3$\,hMpc$^{-1}$.  To test
the robustness of the results we repeated the same analysis reducing
the maximum $k$ fitted $k_{\rm max}$ to 0.2$h/Mpc^{-1}$ obtaining
fully consistent fits, albeit with increased errors. In
method~\ref{method1} we fit simultaneously monopole and quadrupole
(for SGC and NGC with 2 different windows) adopting a 120x120
covariance.  In methods~\ref{method2} and~\ref{method3} we perform a
joint fit of the weighted monopole and quadrupole, $P_{i,w_j} $, each
calculated using the appropriate set of weights for $q_0$ and $q_1$
(and $p_0$, $p_1$); for a 240x240 total covariance including NGC and
SGC samples.

We compute the covariance matrix from the 1000 EZ mocks used in
\citet{2017Mio}, including all weights as,
\begin{equation}  \label{covmat}
C = \frac{1}{999} \sum_{n=1}^{1000}
[\mathbf{d}_n(k_i)-\hat{\mathbf{d}}(k_i)]
[\mathbf{d}_n(k_j)-\hat{\mathbf{d}}(k_j)]^T,
\end{equation}
where $\mathbf{d}_n$ is the vector formed of the multiple weighted
moments being fitted, and $\hat{\mathbf{d}}$ is the mean value. Note
that when inverting the covariance matrix we include the small Hartlap
factor \citep{hartlap2007} to account for the fact that $C$ is
inferred from mock catalogues. An alternative approach would have been
to adjust the Gaussian assumption \citep{2015sellentin}.

Parameter constraints are derived from a MCMC routine, optimised for
this problem. Multiple chains are run for each fit, and convergence is
checked both using the \citet{Gelman92} convergence criteria and by
testing consistency of results from independent chains, starting at
different positions.

\section{Results}  \label{sec:results}

In this section we present the results obtained from the traditional
\ref{method1}, $\Omega_m$ \ref{method2} and $f\sigma_8$ \ref{method3}
analyses; we first present the results obtained assuming a fixed
fiducial distance-redshift relation, i.e. setting $\alpha_\parallel$
and $\alpha_\perp$ both equal to unity in our pipeline
(Section~\ref{fixap}); while in Section~\ref{varap} we allow them to
vary, fitting simultaneously the growth and the geometry.  In
Section~\ref{sec:comparison} we compare the key results of this work
with parallel work performed at a single redshift (as in our
traditional analysis) in configuration space \citep{2017Zarrouk} and
Fourier space \citep{2017Hector}, where the analysis has been extended
to include the hexadecapole moment of the power spectrum. We also compare our results
with the redshift weights based-analysis of \citet{2017Gongbo} in
Section~\ref{sec:comparison}, which makes a number of different
assumptions and explores alternative cosmological models.
 
\textbf{Fiducial cosmology:} we analyse  the data in a flat $\Lambda$CDM cosmological
model with total and baryonic components $\Omega_m(z=0)$= 0.31,  $\Omega_b(z=0)$ = 0.0325;   neutrino masses $\sum m_\nu= 0.06eV$, ampltude of the clustering $\sigma_8(z=0)= 0.8$, spectral index   $n_s= 0.97$ and dimensionless hubble parameter $h= 0.676$; 
\subsection{The weighted multipole measurements}

\begin{figure} \centering
\includegraphics[scale=0.446]{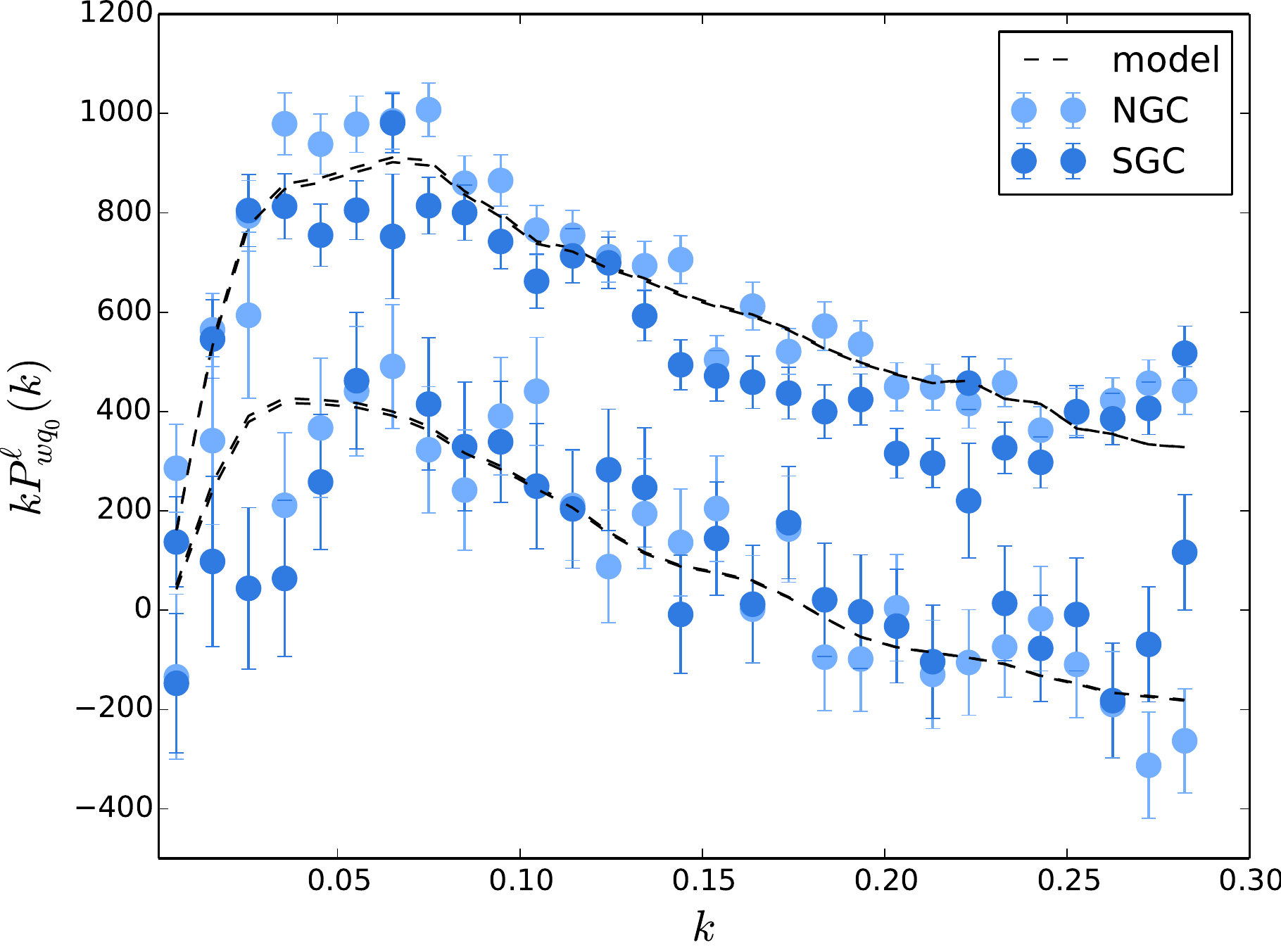}
\caption{The weighted monopole (top) and quadrupole (bottom) for $\Omega_m$ weights;
  we display the measurement of the weighted moments computed using
  the NGC (light blue points) and SGC (blue points) samples. Black
  dashed lines correspond to the best fit models obtained from the
  joint fit of the 2 samples; the 2 best fit dashed lines differ at
  large scales since convolved with two different window functions
  accounting for the different survey geometries and systematics
  between the 2 samples.}
 \label{fig:moments}
\end{figure}
In Fig.~\ref{fig:moments}, we present the moments calculated for the $\Omega_m$
set of weights. They all look very similar for all the weights, showed consistency with
the fiducial $\Lambda$CDM model. It is only if we were to find an
inconsistency with this model, that we would see an anomaly here for a
particular set of weights. i.e. the constraining power lies in the
fact that if the cosmology was very different from the fiducial
$\Lambda$CDM value, these would look very different from each other.
\subsection{Fitting growth in a fixed background geometry}  \label{fixap}

\begin{figure}
\centering
\includegraphics[scale=0.3461]{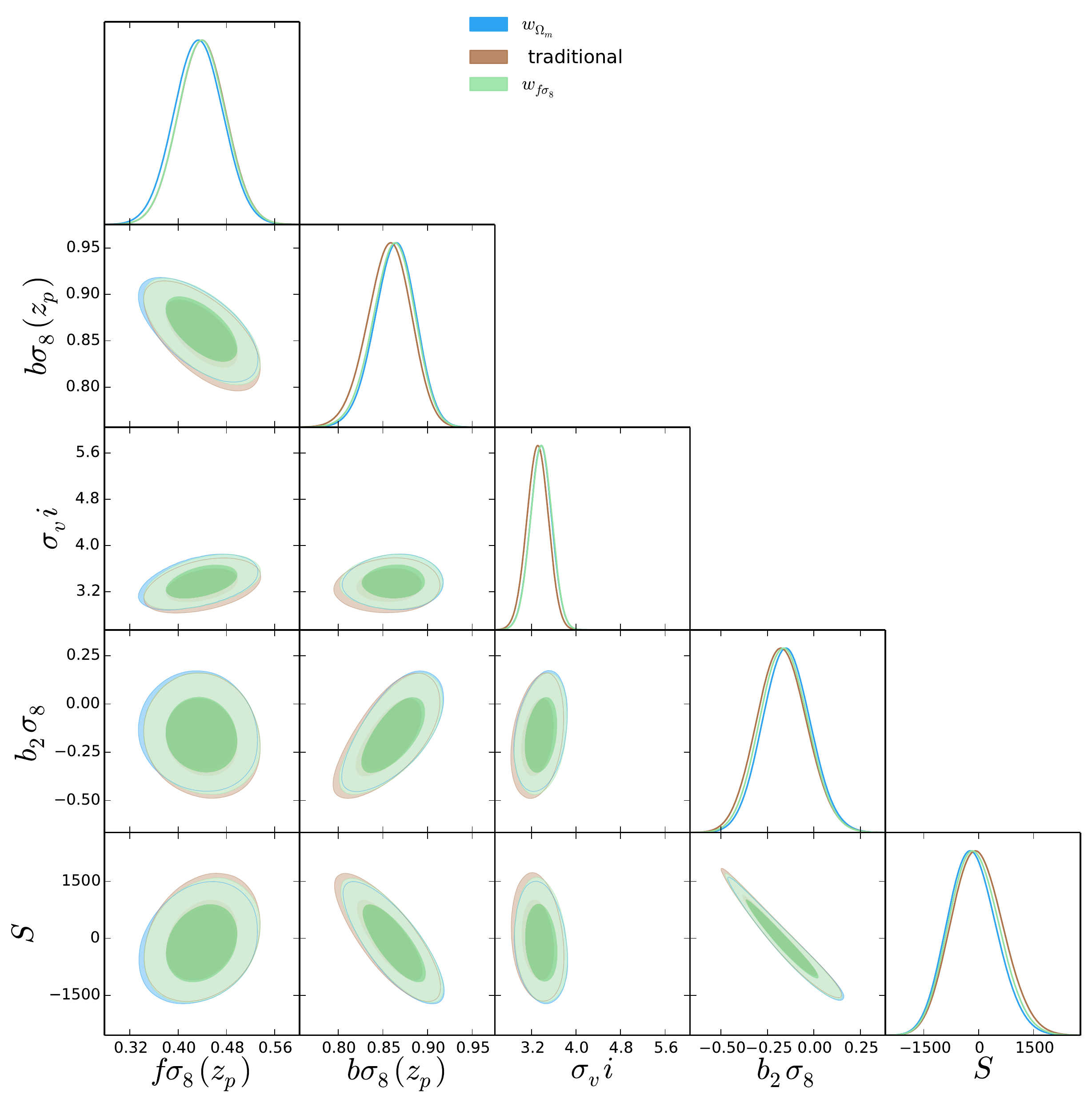}
\caption{A comparison between the values of $f\sigma_8$, $b\sigma_8$,
  $\sigma_{Fog}$, $S$ obtained by the three different methods when the
  background geometry is fixed. Blue and green contours indicates the
  projected values from the $\Omega_m$ and $f\sigma_8 $ analysis (7
  free parameters) respectively; brown contours correspond to the
  constraints obtained from the single-epoch traditional analysis (5
  free parameters).  }
\label{noap1}
\end{figure}

As described in Section \ref{sec:method}, the traditional analysis
constrains the clustering at a single effective epoch allowing for 5
free parameters $f\sigma_8(z_p)$, $b\sigma_8(z_p)$ + nuisance
parameters.  In contrast, the weighted analyses fits the evolution of
$\Omega_m$ and $f\sigma_8$ with redshift, and requires a fit with 7
parameters: $q_0, q_1$ ($p_0$, $p_1$) to model the normalisation and
evolution in the growth, $b\sigma_8(z_p)$
$\partial b\sigma_8/\partial z$ to account for the evolution in the
linear bias $b(z)$, together with nuisance parameters $b_2\sigma_8$,
$\sigma_{\rm FOG}$ and $S$. As we are interested in measuring
cosmological evolution, we need to carefully consider if the nuisance
parameters also need to allow for evolution. Regarding the Fingers-of-God
(FoG), it would theoretically be possible to allow this to vary with
redshift, but we have checked using N-body simulations, that for
$k<0.3$\,hMpc$^{-1}$, the evolution does not impact  $f\sigma_8$; if
we were instead interested in the measurements of non-local bias, for
example, allowing for this evolution would have been a key
requirement. We do allow the bias to be simultaneously fitted as
described in Section~\ref{sec:method}.

In order to compare the redshift-weight measurements with the
traditional one, we projected the 7 parameter MCMC chains
($q_0, q_1, b\sigma_8, \partial b\sigma_8/\partial z$ + nuisance
parameters) into the 5-dimensions parameter space defined at the
effective redshift using the relation $f[\Omega_m(q_0, q_1, z_p )]$,
$f(p_0, p_1, z_p )]$ and consider $b(z)$ at its pivot redshift
value.  The results are displayed in Fig.~\ref{noap1} where we show
likelihood contours for $f\sigma_8(z_p)$, $b\sigma_8(z_p)$, etc as
derived obtained from the three different analysis, traditional (brown
contours), $w_{\Omega_m}$ (blue contours) and $w_{f\sigma_8}$ (green
contours) when imposing $\alpha_\parallel= \alpha_\perp = 1.$ It is
worth noting that all three methods fully agree at the pivot
redshift confirming the previous tests made on the mocks
\citep{2017Mio}. Moreover the redshift weighted analysis give
constraints of the same order as those obtained in the traditional
analysis even though the latter marginalizes over one less free
parameter. This suggests that the information in the data about the
evolution of $f\sigma_8$ is available in addition to the information
obtained at the effective redshift.

\subsection{The  evolution of $f\sigma_8(z)$, $\Omega_m(z)$, $b(z)$ }

\begin{figure} \centering
\includegraphics[scale=0.43]{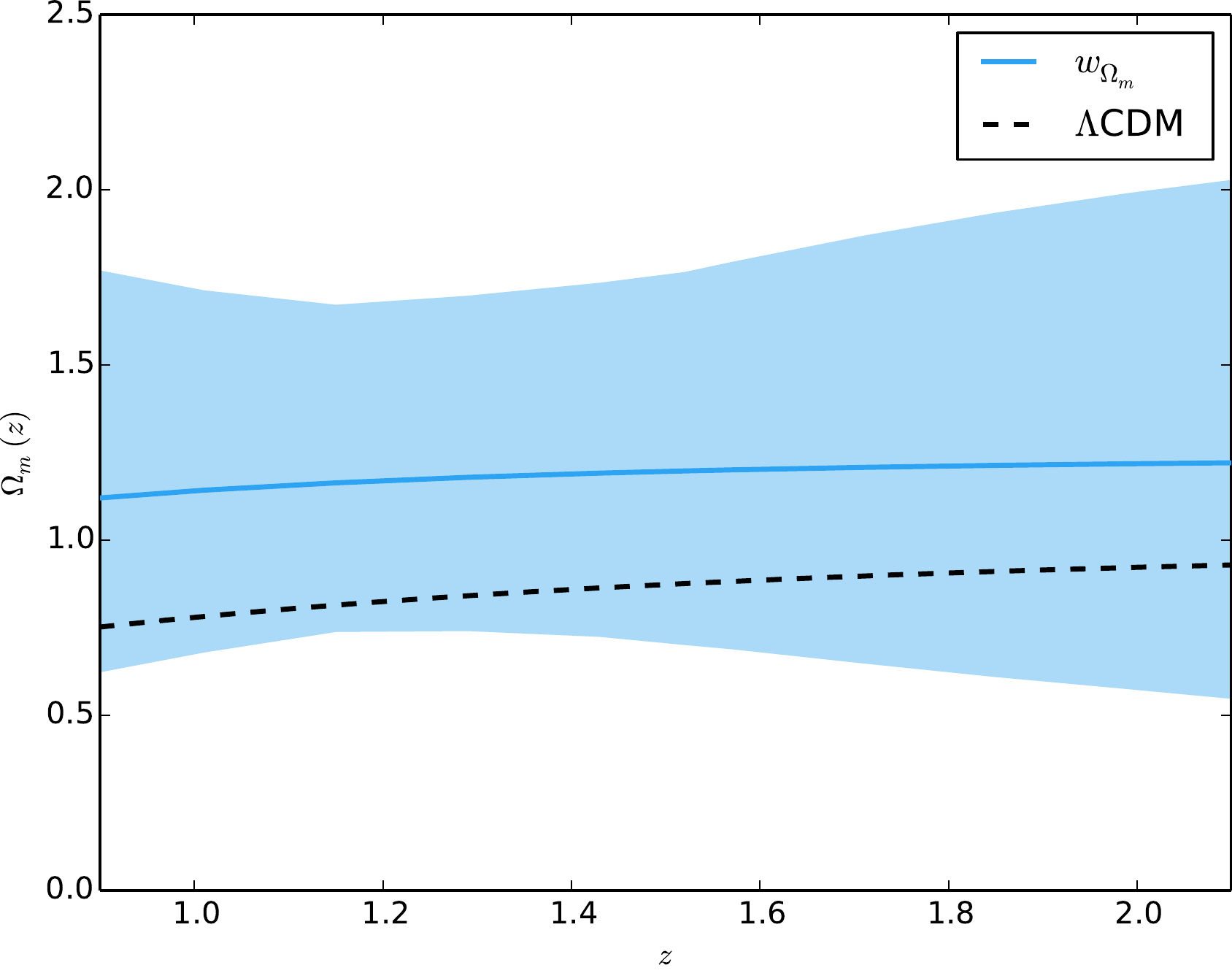}
\includegraphics[scale=0.43]{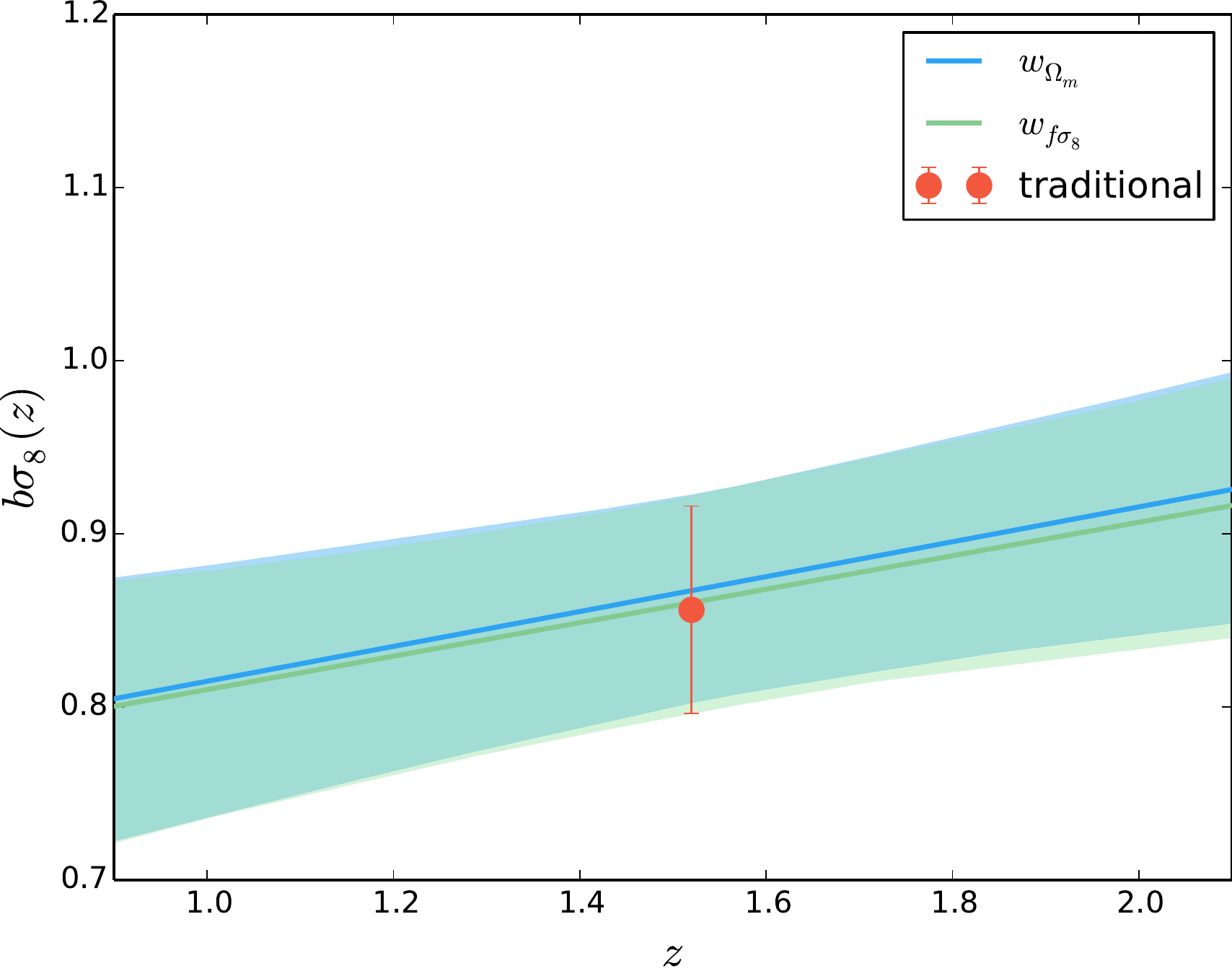}
\includegraphics[scale=0.43]{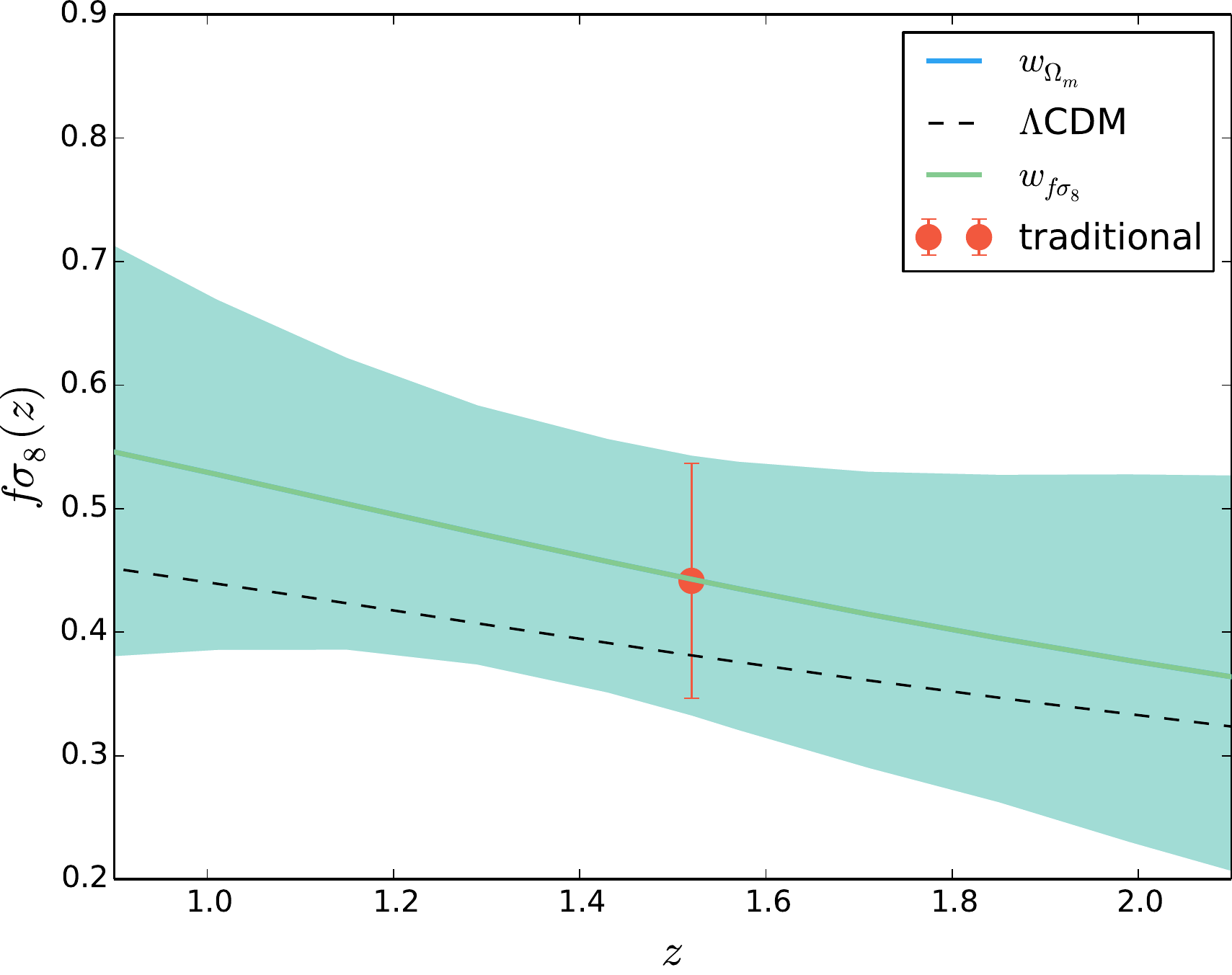}
%\label{7}
%
\caption{Top Panel: the evolution of $\Omega_m(z)$ measured from the
  constraints of $q_0, q_1$. Middle Panel: the evolution of the linear
  bias times $\sigma_8$ fitted using $\Omega_m$ parametrization (blue
  shaded regions), $f\sigma_8$ parametrizations (green shaded
  regions); red point indicates the single-epoch constraints of
  $b\sigma_8(z_p)$ from the traditional analysis.  Bottom Panel: the
  evolution of $f\sigma_8$ from the three different analysis; notation
  and colors as above; all the errors correspond to  68\% confidence level. }
\label{fig:evolution}
\end{figure}

As described and discussed in \citet{2016Mio, 2017Mio}, in general
the redshift-weights allows us to account for the evolution in the
clustering measurements. In this work, through Eq.~(\ref{omz}),
and ~Eq.~(\ref{fsz}) we are able to reconstruct the evolution for
$f\sigma_8$ from both $q_0$, $q_1$ and $p_0$, $p_1$ measurements. We
also modelled a linear evolution of the linear bias as  described by
Eq.~(\ref{bz}).  We show the resulting constraints on the evolution of
$f\sigma_8$, $b(z)$, $\Omega_m(z)$, in Fig.~\ref{fig:evolution}.  

The lower panel of Fig.~\ref{fig:evolution} shows the evolution in
redshift of $f\sigma_8$ obtained applying $w_{\Omega_m}$ (blue shaded
regions) and $w_{f\sigma_8}$ (green shaded regions).  We overplot the
constraints coming from the single epoch (traditional) analysis at
redshift $1.52$. We find good agreement between the different
techniques over the full redshift range. The dashed line indicates the
fiducial cosmology used. We detect a similar slope in the evolution to
that in the fiducial cosmology, and all of our measurement methods
provide results that agree within one sigma with the fiducial
cosmology. The error on $f\sigma_8$ increases while moving from the
pivot redshift in both directions as uncertainties in $q_1$ and $p_1$
become relevant. As we are fixing the projection, varying
$\Omega_m(z)$ only affects the growth rate, explaining the good
agreement between measurements made using both sets of weights: they
are both testing for the same sort of departures from the $\Lambda$CDM
model.

The middle panel of Fig. \ref{fig:evolution} shows our constraints on the linear
redshift evolving bias parameter. Also in this case, we find full
agreement between the different techniques. As mentioned in
Section \ref{sec:method} we do not go beyond linear evolution in the
bias, matching our allowed evolution in the cosmological parameters of
interest. As we are not interested in the recovered bias parameters,
we just want to make sure that the assumptions cannot affect the
constraints we get on the growth rate. \citet{2017Mio} shows that
in this case, the linear assumption is valid.

\subsection{Simultaneously fitting growth and geometry}
\label{varap}

We repeat our analysis, using all three methods, but now including the
projection (AP) parameters in our models. Given the weak detection of
the anisotropic BAO signal in the quasar sample (see \citealt{2017ata}),
a full fit of the monopole and quadrupole is not enough to give
independent strong constraints on the full set of parameters covering
both geometrical and growth-rate deviations. i.e. with only wide
uniform physical priors on the parameters, the degeneracies between
the parameters, particularly the shotnoise term together with
$f\sigma_8$, $b\sigma_8$ $\alpha_\parallel$ and $\alpha_\perp $, does
not allow our chains to converge (after $10^5-10^6$ steps). However, as
pointed out in \cite{2008white}, beyond certain values of
$\alpha_\parallel$ and $\alpha_\perp$, the full background used to
analyse the data loses any meaning. Measurements from independent
cosmological probes in almost all cosmological models that we would
want to test already put tight constraints on these quantities
\citep{2016Planck}. We therefore include a broad prior on both
$\alpha_\parallel$ and $\alpha_\perp$, setting
$0.75 < \alpha_\parallel < 1.25$, $ 0.85 < \alpha_\perp < 1.25$. To
test the robustness of our analysis with respect the choice of the priors we 
performed prior-free analysis  exploring the likelihood surfaces outside of those regions.

\begin{figure}\centering
\includegraphics[scale=0.41]{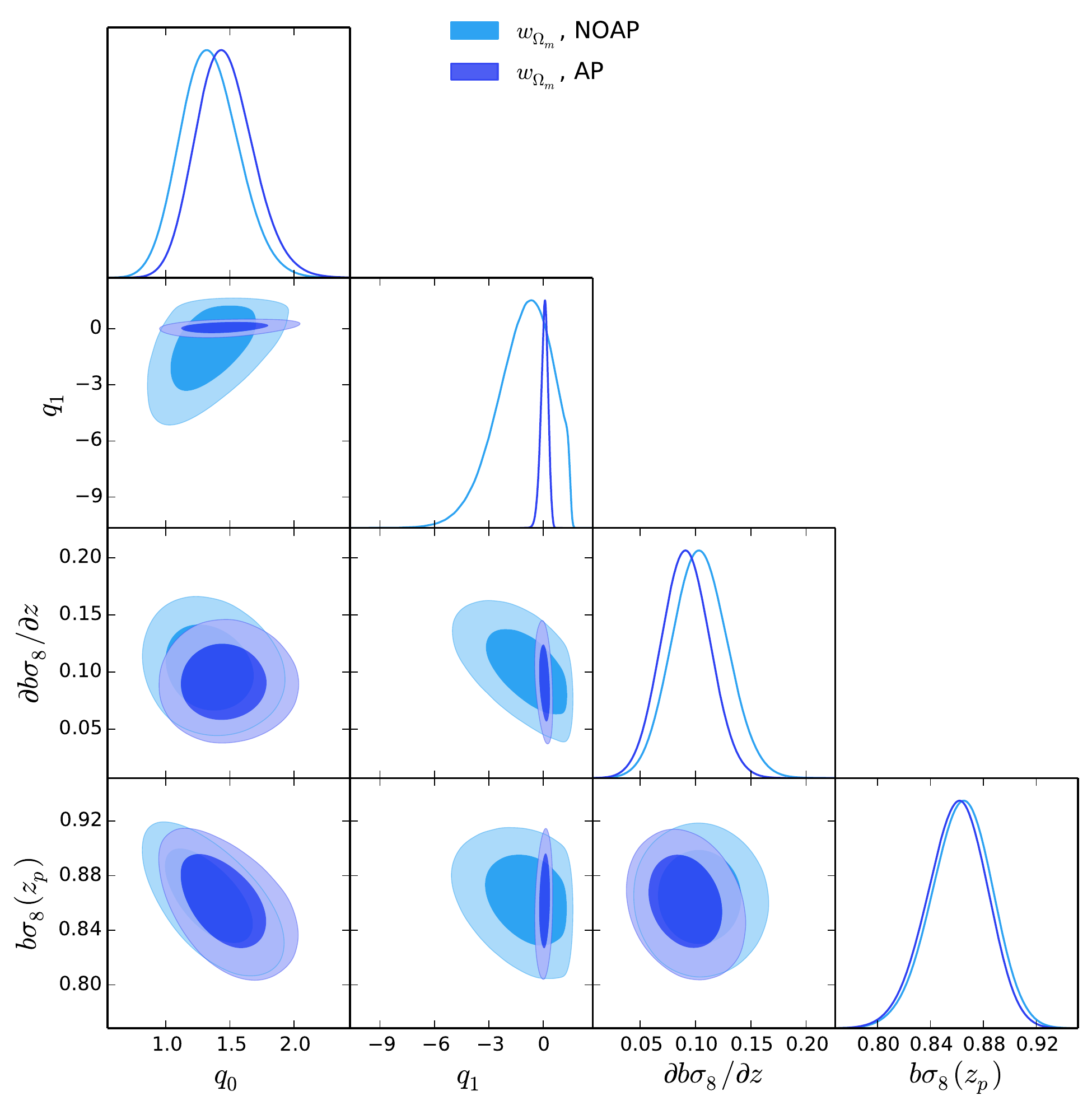}
\caption{A comparison between the constraints on the \textit{evolution parameters} obtained from the $\Omega_m$ analysis with and without fixing the anisotropic projection parameters  (Blue and light blue contours respectively). \label{ompriors} }
\end{figure}

In the traditional and $w_{f \sigma_8}$ analyses we include
$\alpha_\parallel$ and $\alpha_\perp$ as two additional free
parameters. For the $w_{\Omega_m}$ analysis, however, we do not add
any further free parameters: we account for the departures from the
fiducial geometry by including
$\alpha_\parallel[\Omega_m(q_0, q_1 )]$,
$\alpha_\perp[\Omega_m(q_0, q_1 )]$ in our models (as discussed in
Section~\ref{sec:method}). This procedure requires us to impose a
prior on the value of $\Omega_m(z)$ which has to be positive definite
at any redshift to avoid numerical problem; we illustrate the effect
of these prior on the constraints in Fig.~\ref{ompriors}.

\begin{figure}
\includegraphics[scale=0.243675]{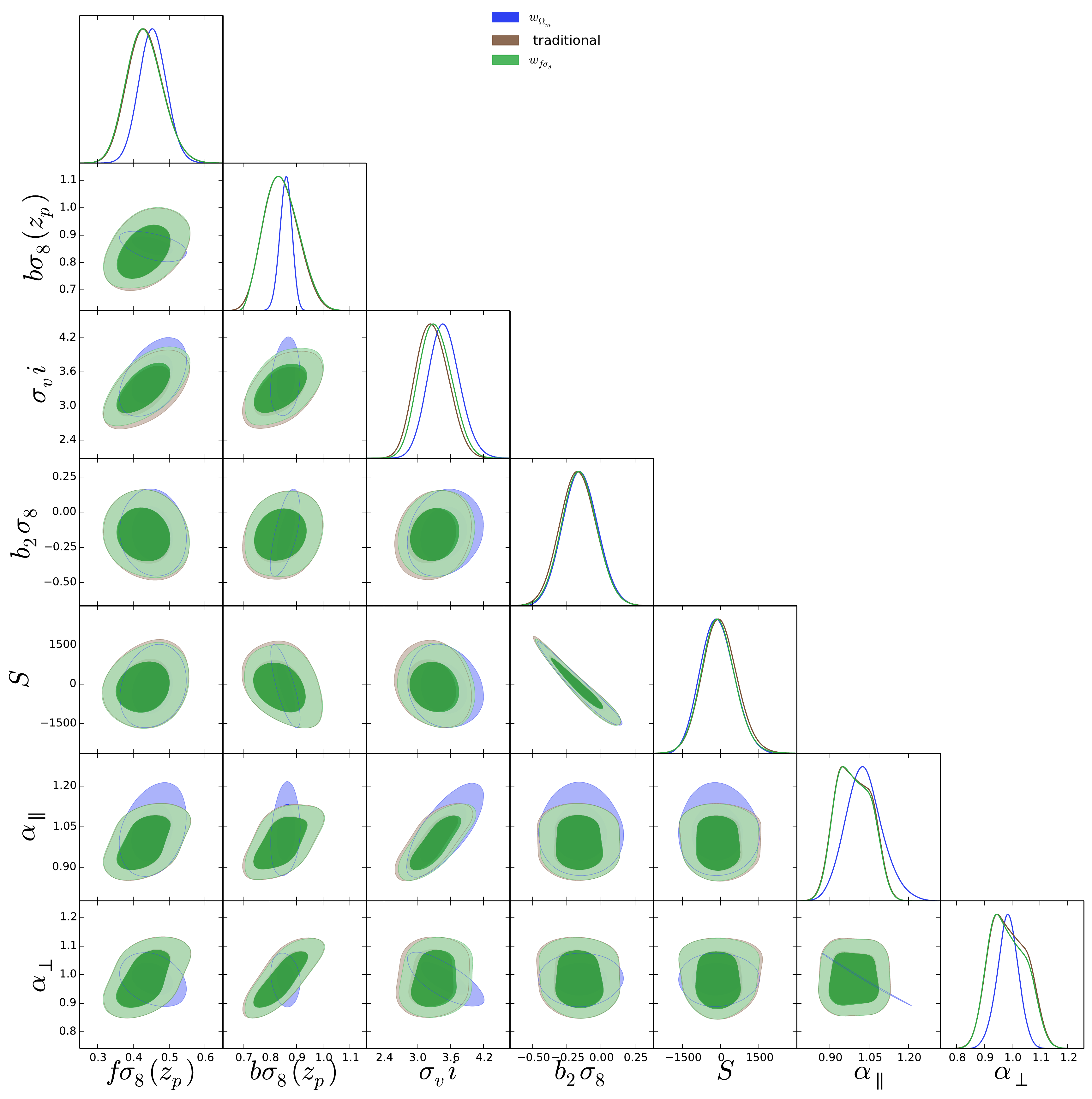}
\caption{A comparison between the values of $f\sigma_8$, $b\sigma_8$, $\sigma_{Fog}$, $S$ obtained from the three different methods (\ref{method1},\ref{method2}, \ref{method3}) when the background geometry is allowed to vary thought the AP parameters. Green contours correspond to the  projected values  from the $f\sigma_8$ analysis ($f\sigma_8(p_0, p_1)$, $b\sigma_8(z_p)$, $\sigma_{Fog}$, $S$, $\alpha_\parallel$, $\alpha_\perp$ ). Blue contours represent the projected constraints from  $\Omega$ analysis  ($f\sigma_8(q_0, q_1)$, $b\sigma_8(z_p)$, $\sigma_{Fog}$, $S$, $\alpha_\parallel(q_0, q_1)$, $\alpha_\perp(q_0, q_1)$ )
%
%and
 %$\Omega_m$ and $f\sigma_8 $ analysis (7 and 9 free parameters) respectively; brown contours %correspond to the constraints obtained from the single-epoch traditional analysis (7 free parameters). 
Brown contours indicate the constraints from the single-epoch traditional analysis at the pivot redshift $z =1.52$  }
\label{apmet123}
\end{figure}

\begin{figure}\centering
\includegraphics[scale=0.41]{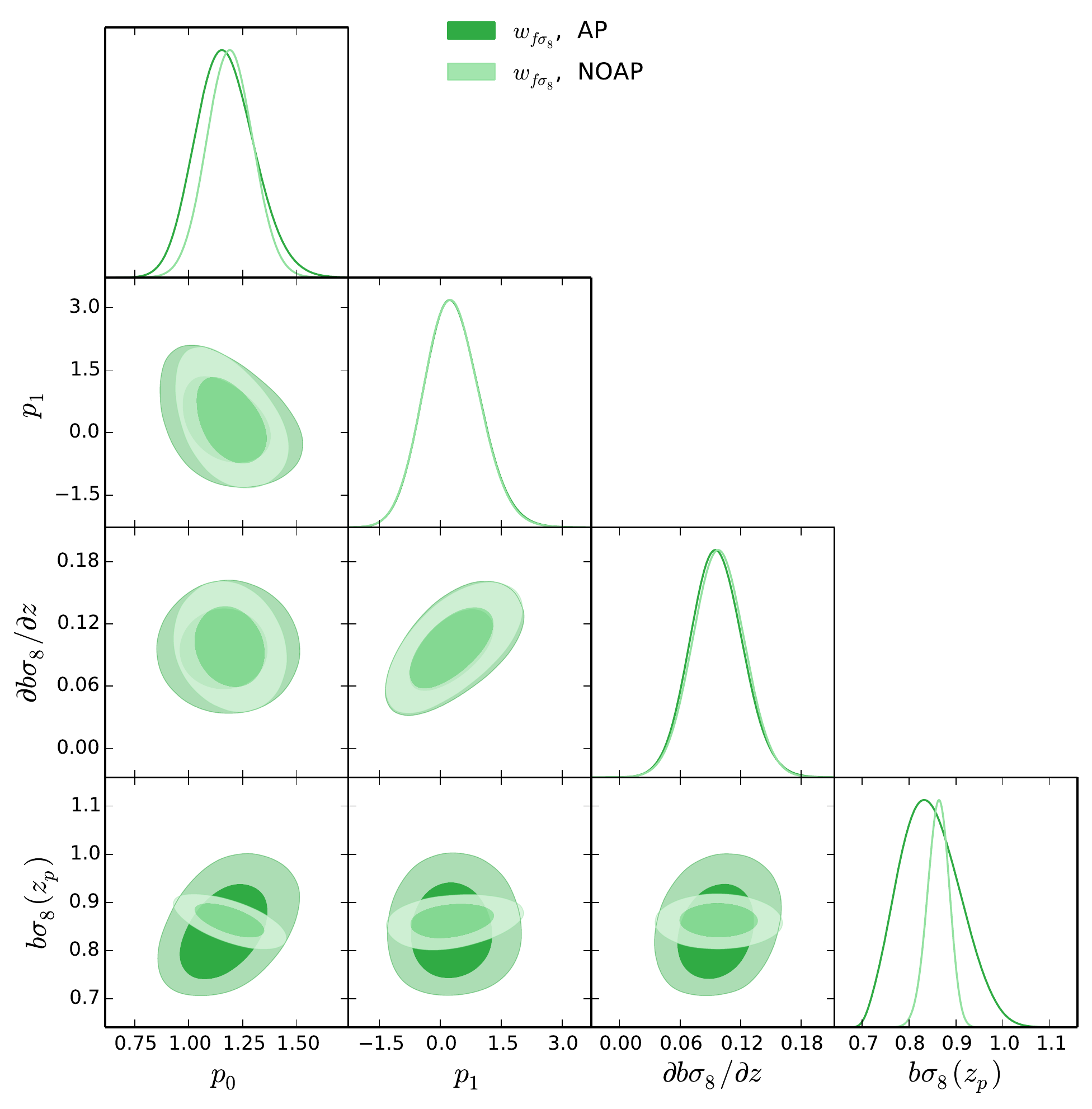}
\caption{A comparison between the constraints on the \textit{evolution parameters} obtained from the $f\sigma_8$ analysis with and without fixing the anisotropic projection parameters  (Green and light green contours respectively). }
\label{fs8priors}
\end{figure}

\begin{figure}
\includegraphics[scale=0.6]{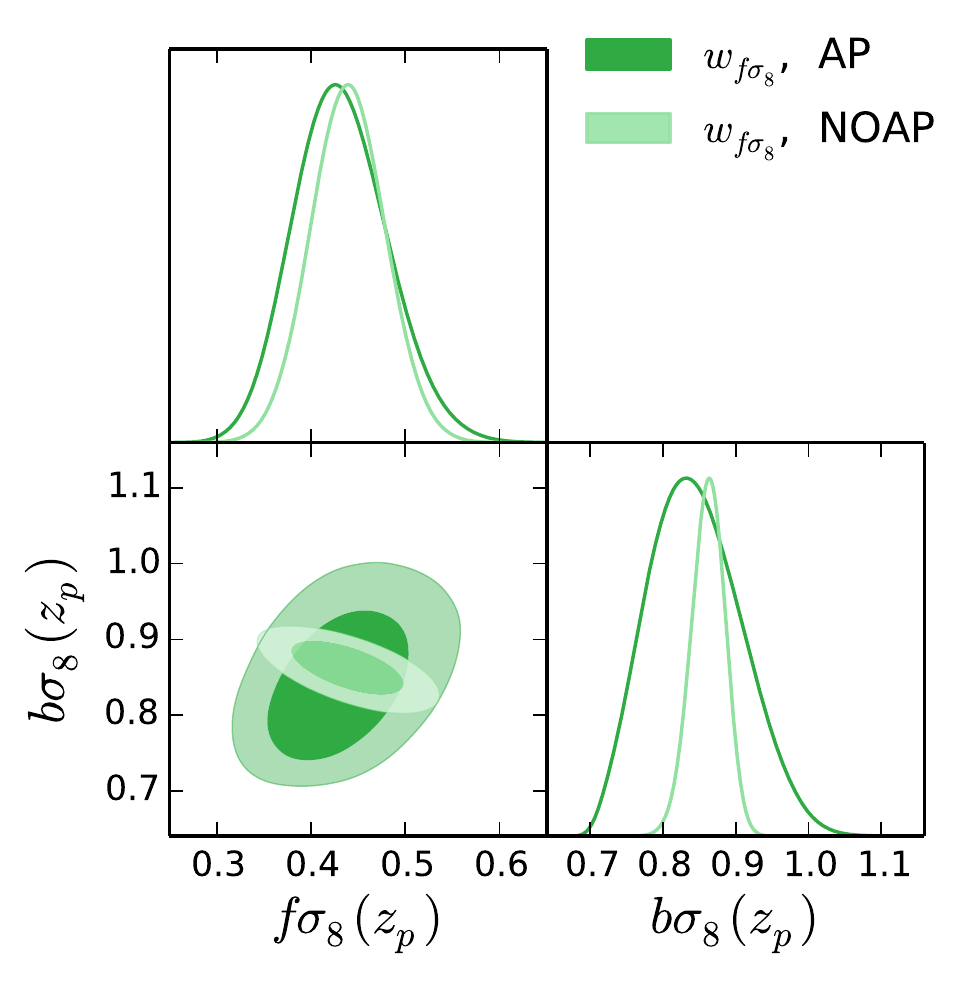}
\includegraphics[scale=0.6]{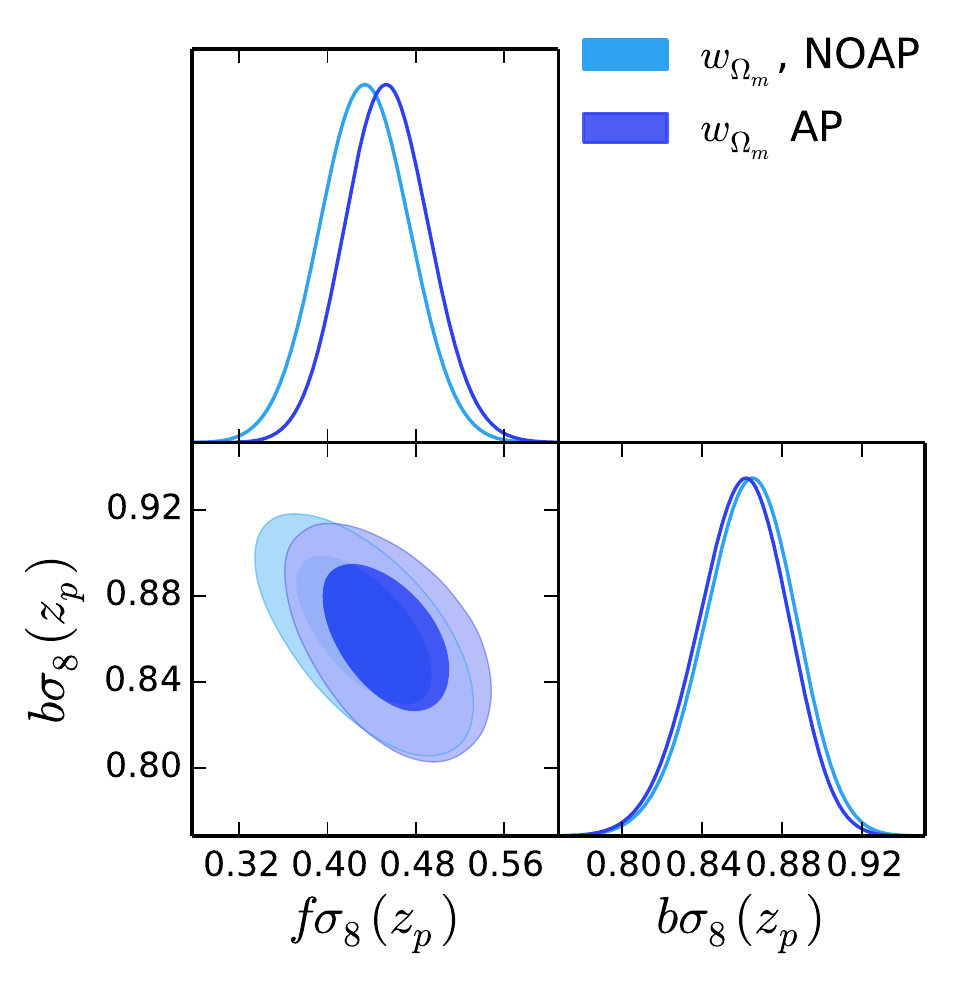}
\includegraphics[scale=0.6]{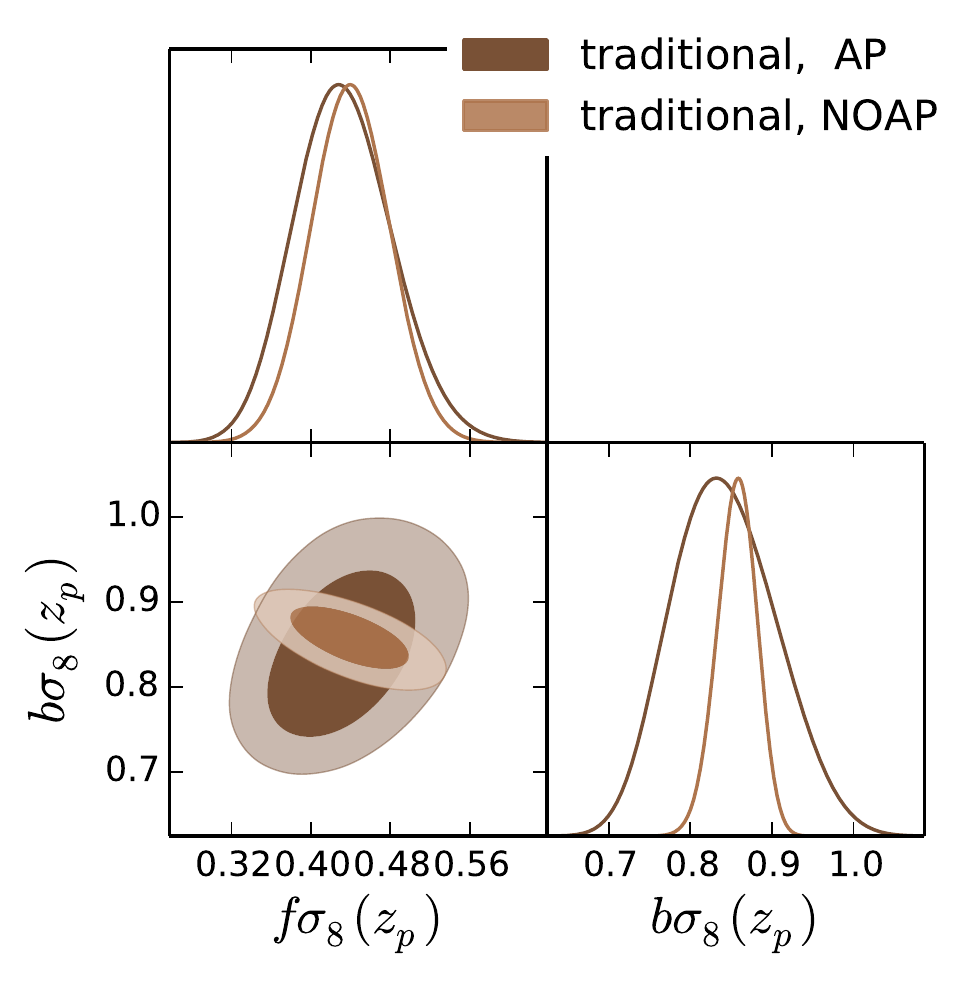}
\caption{ A comparison between the different analysis with AP (darker colors) and without AP (lighter colors). 
Bottom panel shows the constraints from the traditional analysis of $f\sigma_8(z_p)$ and $b\sigma_8(z_p)$.  
Middle and top panel for the projected constraints of $f\sigma_8(q_0, q_1), b\sigma_8(q_0, q_1)$ (blue and light blue contours) and
$f\sigma_8(p_0, p_1), b\sigma_8(p_0, p_1)$ (green and dark green contours)  in the  $w_{f\sigma_8}$ $w_{\Omega_m}$ analysis. 
%on the \textit{evolution parameters} obtained from the $\Omega_m$ analysis with and without %APeffect. (Blue and light blue contours respectively. 
 }
\label{q01apnoap}
\end{figure}

Fig. \ref{apmet123} shows the likelihood contours obtained from the
three different analysis when allowing for unknown projection
parameters (the $AP$ parameters). Dark brown contours refer to
\ref{method1}; The constraints for the $w_{\Omega_m}, w_{f\sigma_8}$
analysis (dark blu $w_{\Omega_m, AP}$ and dark green,  $w_{f\sigma_8, AP}$) are obtained projecting $q_0, q_1$
and $p_0, p_1$ into $f\sigma_8(q_i)$ $(p_i)$; also in this scenario we
confirm a good agreement between the three analyses; as explained,
$\alpha_\parallel$ and $\alpha_\perp$ are not free in the
$w_{\Omega_m}$ analysis but we derive them from the constraints of
$q_0, q_1$. This is the reason why the two parameters are highly
correlated, as shown in the Figure. Note that the $w_{\Omega_m}$ method has two
less free parameters with respect $w_{f \sigma_8}$ and one less 
 with respect the traditional analysis.  In
Figure ~\ref{ompriors} we compare the \textit{evolution} parameters
$q_{0,1}, b\sigma_8, \partial b\sigma_8/\partial z|_{z_p} $ obtained
(dark blue contours,  $w_{\Omega_m, AP}$), with previous results when the geometry has been fixed
 (blue contours, $w_{\Omega_m, NOAP}$).  We find a good
agreement between the two; the shapes of $q_1$ likelihoods show the
effect of the physical priors we are including: $\Omega_m(z_p) >0 $ for
$w_{\Omega_m \rm NOAP}$ and $\Omega_m(z) >0, \; 0.0 < z < 2.2 $ for
$w_{\Omega_m \rm AP}$.  Fig.~\ref{fs8priors} is structured in the same
way as Fig.~\ref{ompriors}; we compare the results from $w_{f\sigma_8, AP}$  with
previous results of $w_{f\sigma_8, NOAP}$ method. We find a good agreement with
the best fit values obtained; note that here we do not assume physical
priors on the sign of $\Omega_m$.  Finally in Fig. ~\ref{q01apnoap} we
compare the constraints at the pivot redshift for $f\sigma_8$ and
$b\sigma_8$ with and without AP, for method~\ref{method1},
\ref{method2} and~\ref{method3} (brown, blue, green contours); we confirm the good agreement on the constraints for $f \sigma_8$ with and withouth fixing the geometry. When  performing the anisotropic fit we get a larger  error as expected; note that for the $w_{\Omega_m}$ analysis we get the constraints to be of the same order: as  explained, in this scenario, we  tie together geometry and growth, thus $\alpha_\parallel$ and $\alpha_\perp$ are not independent parameters.

%	Possible implementations/more discussion to be included:
%	\begin{itemize}
%	%\item roughly the same description as above of the different plots 
%	%\item include discussion on the flat prior (giving details of those)
%   on $\alpha_\parallel $ and $\alpha_\perp$: this priors are necessary given the signal
%   in the data and still much larger than planck + quote martin white paper (2008). 
%	
%	\item  Relax the priors including hexadecapole; in principle when doing RSD weights 
%   techinque we are not interested in this since  $P_4$ is highly affected from window + our
%   ignorance about non linearities and their evolution in redshift. \ref{Hector and Pauline}. 
%	\item Fully consistent results between ap and noap; different priors in the $\Omega_m$ techinque
%	% has a  case require a particular explanation (same number of parameters, etc)
%	\item  bs8 vs  AP.
%	\end{itemize}
%	 In Fig.~\ref{noap0a}, and
%	Fig.~\ref{noap0b} we show the likelihood contours for
%	$q_0, q_1, b\sigma_8(z_p) , \partial b\sigma_8/\partial z|_{z_p}$ and $p_0, p_1,
%   b\sigma_8(z_p), \partial b\sigma_8/\partial z|_{z_p}$
%	respectively. We overplot the expected values for $q_0$ and
%	$q_1$ (and $p_0$, $p_1$) showing that the results are fully consistent
%	with $\Lambda$CDM. 

\begin{figure}\centering
\includegraphics[scale=0.42]{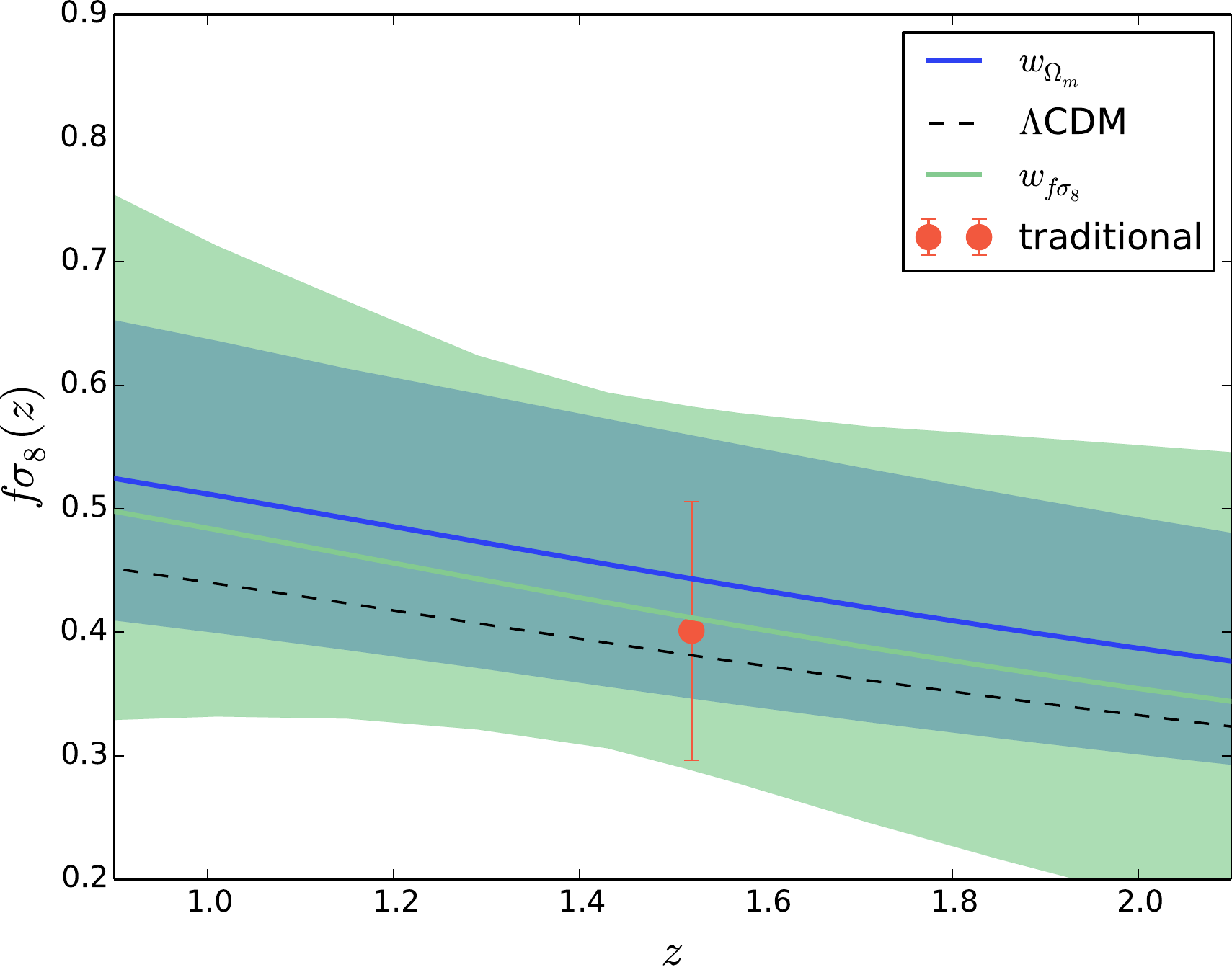}
\includegraphics[scale=0.42]{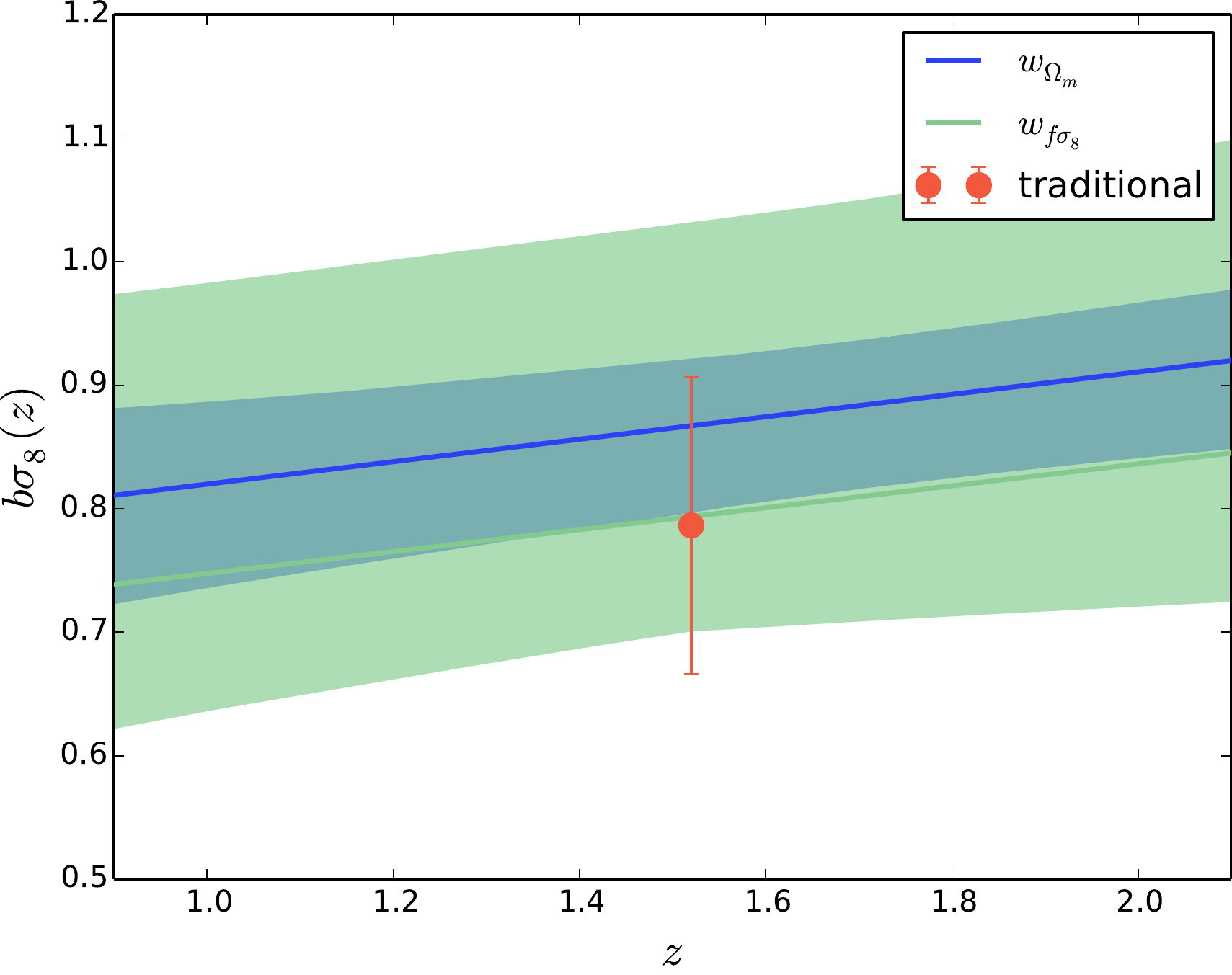}
\caption{ The evolution of $f\sigma_8(z)$ and $b\sigma_8(z)$ when including the AP effect; notation and colors same as in \ref{fig:evolution};  all the errors correspond to  68\% confidence level. }
\label{fig:evolution2}
\end{figure}

\subsection{Bestfit measurements}
In Table~\ref{table1} we summarize the results from the different analysis (\ref{method1}, \ref{method2}, \ref{method3}) with and without free AP parameters (bottom and top panel). 
We display the best fit values (first column)  the mean values $\pm 1\sigma $ (second column); 
  The first section of the table shows the
    fit to the monopole and  quadrupole fixing the AP parameters. While
    the second section of the table shows the fit results allowing the
    AP parameters to be simultaneously fitted.  The fitting range is
    $k = 0.01$ - $0.3\ihMpc$ for both the monopole and quadrupole.
    We consider the results from combining both North Galactic Cap
    (NGC) and South Galactic Cap (SGC) using standard redshifts
    estimates. The error-bars are obtained by marginalising over all
    other parameters.

\begin{table*}\label{table1}
\begin{center}
  \caption{The best fitting measurements for the DR14 quasar data over the redshift range $( 0.8 <z< 2.2)$.
 Left panel for results with fixed anisotropic projection parameters (NOAP). Right panel  for results with free anisotropic projection parameters (AP).     
 These are the marginalised constraints made from the chains presented in Figs \ref{fig:evolution}, \ref{fs8priors} and \ref{fig:evolution2}.}

	\begin{tabular}{l l l l }
%\toprule
%\rule{2cm}{0.4pt}
&&NOAP\\
%%%%%STANDARD
\midrule
           $Traditional $ \\
 &max. like.& mean $\pm 1\sigma$ \\
		$f(z)\sigma_{8}(z)$    &    0.435  &  0.44   & $\pm $   0.04  \\
		$b\sigma_8          $ 		     &    0.86   &  0.86   & $\pm $   0.02  \\
		$\sigma_{FOG}   $        &    3.30   &  3.30   & $\pm $   0.19  \\
		$b_2\sigma_8     $	       &   -0.18   & -0.17   & $\pm $   0.13  \\
		$ S  $ 				   &   -340    &  -270   & $\pm $    697  \\
		$\chi^2 $ 									 & 113/(120 - 5)\\ %  & 115/(120 - 5)  \\ 
\hdashline
%%%%OMEGA	
$\Omega_m$ weights\\
%\midrule
		$q_0$  		                                 &  1.31    &  1.34 & $\pm$ 0.23    \\
		$q_1$ 		                                 & -1.07    & -1.09 & $\pm$ 1.50    \\
		$b\sigma_8(z_p)$ 	                         &  0.10    &  0.10 & $\pm$ 0.025   \\
		$\partial b\sigma_8 /\partial z |_{z_p}$ 	 &  0.87    &  0.86 & $\pm$ 0.02    \\
		$\sigma_{FOG} $								 &  3.39    &  3.34 & $\pm$ 0.19    \\
		$b_2\sigma_8  $    							 & -0.15    & -0.15 & $\pm$ 0.13    \\
		$ S  $										 & -208     & -174  & $\pm$ 660  	  \\
		$\chi^2 $ 									 & 221/(240 - 7) \\%&     223/(240 - 7)  \\ 

%\midrule
%%%%%fs8	
\hdashline
$f \sigma_8$ weights\\
%\midrule
		$p_0$  		                                 & 1.11   &   1.12 &  $\pm$ 0.11   \\
		$p_1$ 		                                 & 0.35   &   0.28 &  $\pm$ 0.69   \\
		$b\sigma_8(z_p)$ 	                         & 0.865  &   0.86 &  $\pm$ 0.02   \\
		$\partial b\sigma_8 /\partial z |_{z_p}$ 	 & 0.10   &   0.10 &  $\pm$ 0.03   \\
		$\sigma_{FOG} $								 & 3.33   &   3.37 &  $\pm$ 0.19   \\
		$b_2\sigma_8  $    							 &-0.15   &  -0.16 &  $\pm$ 0.13   \\
		$ S  $									     & -218   &  -106  &  $\pm$  676   \\ 
		$\chi^2 $ 									 & 223/(240 - 7) \\% &  225/(240 - 7)  \\ 
%\midrule
%
%
%
			  \end{tabular}
			  \begin{tabular}{l l l l }
%
%	
%\midrule
&& AP\\
\midrule
 $Traditional $\\
 &max. like.& mean $\pm 1\sigma$ \\
        $f(z)\sigma_{8}(z)$  & 0.40   & 0.43   &  $\pm$ 0.05      \\
		$b\sigma_8$ 		 & 0.79   & 0.84   &  $\pm$ 0.06      \\
		$\sigma_{FOG} $		 & 3.0    &  3.2   &  $\pm$ 0.29      \\
		$b_2\sigma_8  $    	 &-0.16   & -0.17  &  $\pm$ 0.13      \\
		$ S  $				 & 28     &  -37   &  $\pm$ 685       \\		
		$\alpha_{\parallel}$ & 0.95   &  0.99  &  $\pm$ 0.065     \\
		$\alpha_{\perp}$     & 0.94   &  0.99  &  $\pm$ 0.06      \\
		$\chi^2 $ 			 & 112/(120 - 7)  \\%  & 114/(120 - 7)  \\ 
\hdashline
$\Omega_m$ weights\\        %
		$q_0$  									  &  1.42     & 1.46  &  $\pm $ 0.22    \\
		$q_1$ 									  &  0.07     & 0.07  &  $\pm $ 0.20    \\
		$b\sigma_8(z_p)$ 						  &  0.86     & 0.09  &  $\pm $ 0.02   \\
		$\partial b\sigma_8 /\partial z |_{z_p}$  &  0.09     & 0.09  &  $\pm $ 0.02    \\
		$\sigma_{FOG} $							  &  3.44     & 3.48  &  $\pm $ 0.28    \\
		$b_2\sigma_8  $    						  &  -0.16    & -0.15 &  $\pm $ 0.13    \\
		$ S  $									  &  -145     & -124  &  $\pm $  653    \\
		$\chi^2 $ 								  & 222/(240 - 7)  \\%   &  224/(240 - 7)  \\ 

\hdashline
$f \sigma_8$ weights\\
		$p_0$  		                                 &  1.11  &  1.11  &  $\pm$ 0.13    \\
		$p_1$ 		                                 &  0.16  &  0.29  &  $\pm$ 0.69    \\
		$b\sigma_8(z_p)$ 							 &  0.79  &  0.85  &  $\pm$ 0.06    \\
		$\partial b\sigma_8 /\partial z |_{z_p}$ 	 &  0.09  &  0.10  &  $\pm$ 0.03    \\
		$\sigma_{FOG} $								 &  3.19  &  3.33  &  $\pm$ 0.29    \\
		$b_2\sigma_8  $    							 & -0.13  & -0.16  &  $\pm$ 0.13    \\
		$ S  $									 	 & -205   & -95    &  $\pm$ 664  	 \\
		$\alpha_{\parallel}$ 						 &  0.94  &  0.99  &  $\pm$ 0.06    \\
		$\alpha_{\perp}$   						     &  0.94  &  0.98  &  $\pm$ 0.06    \\
		$\chi^2 $ 									 & 222/(240 - 9)  \\%&   224/(240 - 9)  \\
\bottomrule
 \end{tabular}
	  \label{tab:results}
\end{center}
\end{table*}
\subsection{Consensus with other projects}  \label{sec:comparison}
\begin{figure}\centering
\includegraphics[scale=0.42]{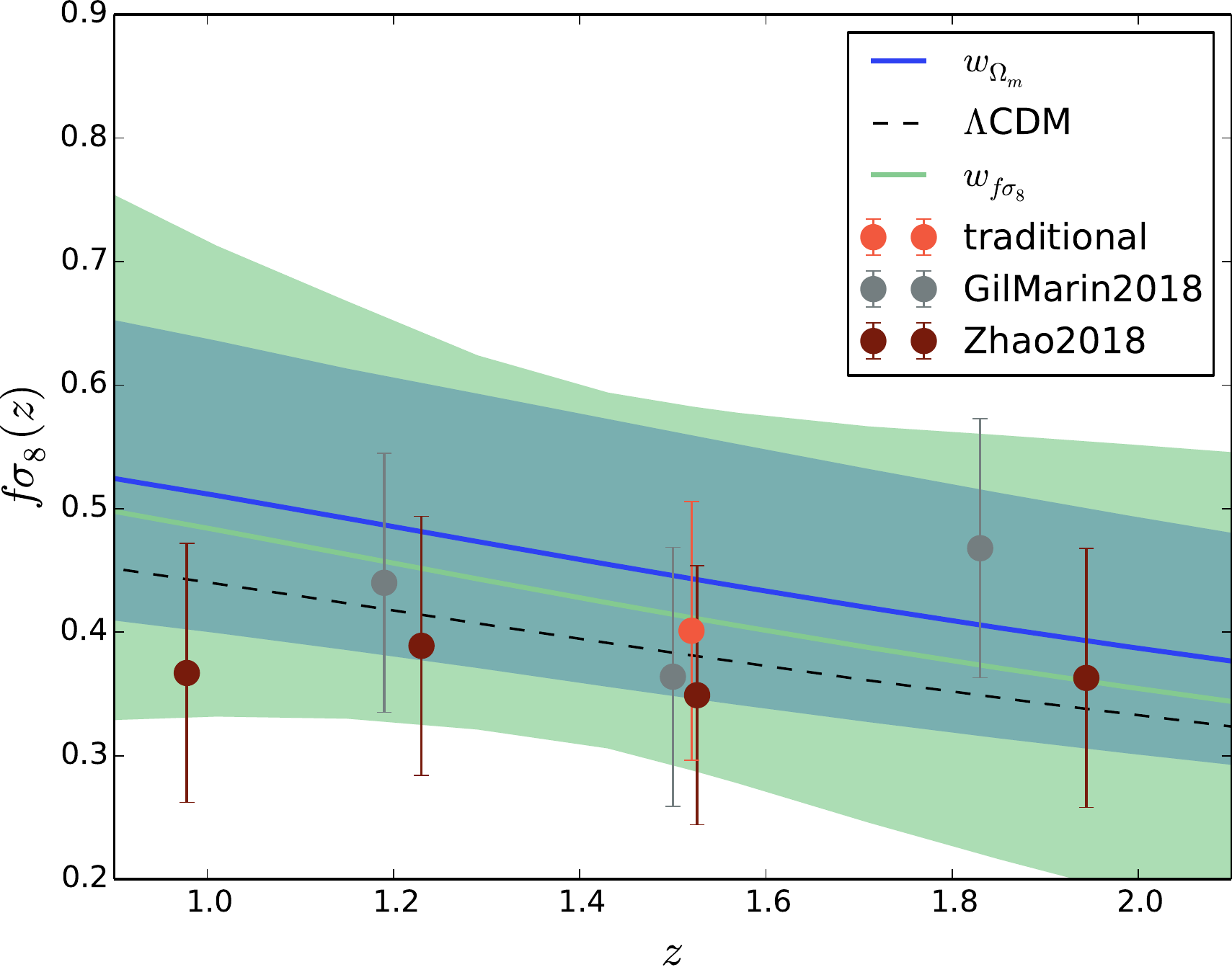}
\caption{comparison of $f\sigma_8$ evolution as obtained by different analysis.  All the errors correspond to  68\% confidence level.
%: band in z, fs8 measurements: 3 point of hector, 4 point for gongbo, traditional for everyone
\label{fcomp}
}
\end{figure}

The current analysis has been compared with similar analysis performed on the same data set
\citep{2017Zarrouk, 2017Hou, 2017Gongbo, 2017Hector}; 
we refer to \cite{2017Zarrouk} for a  longer discussion on the different methodologies and we here focus on the comparison only between  analyses  measuring the redshift evolution of the growth rate.
% clustering. 
In particular we compare our results with analyses presented in \cite{2017Hector} and \cite{2017Gongbo}. 
In \cite{2017Hector} the evolution of $f\sigma_8$ have been studied performing an analysis in three different overlapping redshift bins: $0.8 <z< 1.5$,  $1.2 <z< 1.8$,  $1.5  <z< 2.2$, corresponding to effective redshifts $1.19; 1.5; 1.83; $
This standard analysis considers the first three moments of the power spectrum, $P_{0,2,4}$, up to $k = 0.3 h\rm Mpc^{-1}$; the measurements are fitted with the TNS model computed up to 2-loop in standard perturbation theory; the  window survey effect is accounted following  \cite{2017wilson}. 
In \cite{2017Gongbo} they perform a  joint BAO and RSD analysis  using the monopole and quadrupole (in the $k$-range of $0.02\leq k\,[h{\rm Mpc}^{-1}]\leq 0.30$) and comparing with a TNS redshift space power spectrum template at 2-loop level in perturbation theory;
 They derive redshift weights following the lines of \citep{zhu2014, 2016Mio} to optimize the constraints on  $\alpha_{\bot}, \alpha_{\|}$ and $f\sigma_8$ at four effective redshifts, namely, $z_{\rm eff}=0.98, 1.23, 1.53$ and $1.94$.
In contrast to the analysis presented in this work where  the whole redshift range is considered and the  weighted multipoles are combined in a joint fit, in \cite{2017Gongbo} the redshift weights act to divide the sample into \textit{smooth} z-bins. In each bin they perform the same analysis to constrain  $f\sigma_8(z_{\rm eff}), \alpha_{\parallel, \perp}(z_{\rm eff})$ at the four effective redshifts. Thus this approach is a hybrid between redshift-weighting and standard analyses. \cite{2017Gongbo} use  an optimisation to find the best redshift kernels and then perform a standard analysis for each, assuming the measurements being at one effective redshift. In contrast we directly measure parameters controlling the redshift evolution. 
In Figure \ref{fcomp} we show  the constraints from the different analysis.The
red point and blue and green band correspond to the traditional and redshift  weigth analysis results \ref{method1}, \ref{method2}, \ref{method3} presented in this work. Grey points correspond to the redshift bin analysis presented in  \cite{2017Hector}, while dark red points correspond to the analysis of \cite{2017Gongbo}. We confirm the good agreement between the different techniques in measuring $f\sigma_8(z)$. Note that the marginalized error bars for the red, grey, dark red points refer to analyses with 7 free parameters  ($f\sigma_8 (z_{eff}), b\sigma_8(z_{eff}), \alpha_\parallel, \alpha_\perp +\rm  nuisance$) while redshift weights methods (blue and green band) include 7 and 9 free parameters respectively ($q_0, q_1, b\sigma_8, \partial b\sigma_8/\partial z + nuisance$),  ($p_0, p_1, b\sigma_8, \partial b\sigma_8/\partial z \alpha_\parallel, \alpha_\perp , + nuisance$). 
%It includes, (To be updated soon)
%
%	\begin{itemize}
%	%
%	\item list of the measurements from the other papers
%	\item  number of free parameters/statistics used in the different analysis. %, $chi^2$
%	%\item likelihood surfaces and different correlations (comparing with Gongbo, Hector)
%	\item redshift weights vs redshift binning (comparing with Hector)
%	\end{itemize}
%
\section{ discussion}  \label{sec:discussion}
The DR14 quasar sample  allows for tests of the cosmological model at previously unexplored epochs; further, as   it  also  cover  a wide redshift range $0.8-2.2$, it 
opens up the possibility of directly investigating the evolution of the cosmological parameters. % in the full volume considered. 
Standard analyses (e.g. \citealt{Alam2016})  investigate the evolution of the growth rate at different epochs by cutting their volume into  redshifts slices.
The quasar sample is characterized by a low density compared to previous samples, such as the BOSS LRG sample, thus  the bin-cutting can have a significant impact on the resulting S/N. 

In this work, we choose to constrain the growth rate and its first derivative in redshift 
considering the full redshift range, using optimal redshift weighting techniques; redshift weights act as a smooth window on the data,  compressing the correlation in the redshift direction, while keeping track of the underlying evolution of the clustering.
We select the optimal redshift weights as they are predicted through the Fisher matrix. The weights specific for the growth measurements have been  derived in \cite{2016Mio} and tested in \cite{2017Mio}. 

We explore two different parametrizations to model the evolution in redshift of $f\sigma_8$; the first models the evolution in redshift through $\Omega_m$(z). This parametrization allows us to account simultaneously for deviations  in both geometry and growth  with respect to the $\Lambda$CDM scenario. 
 The second,  investigates deviations in the evolution of $f\sigma_8(z)$ about the fiducial cosmology; in this case the growth and the geometry deviations are artificially kept separated.

To compare the  constraints on $f\sigma_8$ with the \textit{traditional} method, performed at a single epoch,  we computed  $f\sigma_8(z_p)$ from the evolving constraints, finding full agreement between the three different methods. 

We perform the same analysis first by fixing  the geometrical projection, given by $H$, $D_A$: in this case as expected both redshift weight methods give exactly the same constraints of $f\sigma_8$. We then considered an anisotropic fit, including the AP parameters in our models. In this case the constraints  from
$\Omega_m(z) $ differ with the other analyisis since $\alpha_{\parallel,\perp}$ are not included as free independent parameters but their evolution is described through  $\Omega_m(z)$. Also in this scenario we find good agreement (within $1 \sigma$) between the parameters of interest. 

In this and in  \cite{2016Mio,2017Mio} we showed step by step how to include the redshift weights in the analysis; we also showed how easily to account for the evolution in the models by re-deriving the window function and confirmed that the redshift weights method gives unbiased constraints.  Future surveys  are expected to reduce the statistical error by an order of magnitude over a wide redshift range.
 Therefore, it will be  be increasingly important to account for the evolution in the models.
The extent of the dynamical redshift range covered, by for example DESI  \citep{DESREF} will open up the possibility to discriminate between different cosmological scenarios.  This will be accomplished using the evolution of the key-parameters to remove part of the degeneracy between them. 

%\textbf{some example? ELG redshift ranges, prediction for errors. }

%	%\item  why quasars are
%	\item r
%	\item We account for the evolution in redshift in our models thought 2 different parametrization: one with respect to $ \Omega_m$ which can account for departure of different nature from LCDM, at the same time; as long as those deviations are small enough so that we can assume still valid the relation between growth and geometry w.r.t omegam.
%	% of deviation at the same time,  ( as long as close to LCDM), the other one to explore larger deviations and able to interpret those. The other one correspond to a more standard rsd measurements that keep geometry and growth separated.  We also consider a Standard analysis to compare with. 
%	% which fully agrees at the single epoch; 
%	We performed the same three analysis in the case of $\alpha_\parallel$ and $\alpha_\perp$ fixed and able to vary. 
%	We found agreement within one sigma with LCDM. 
%\item what we presented: 3 different analysis all consistent between them 
%
%
%\item say something like: very easy to include the weights and explore different %optimization/evolution.
%
%we hope to be able to model and capture non linearities accurately.
%\item what to expect from the same techinque (rsd weights) when measuring better surveys future. 
%\end{itemize}

\section*{Acknowledgements}

RR and WJP acknowledge support from the European Research Council
through the Darksurvey grant 614030. 
RR also thanks Dr. Violeta Gonzalez-Perez, Dr. Hans Winther, Dr. Seshadri Nadathur,  Iza Pstrucha and Gary Burton.
WJP also acknowledges support
from the UK Science and Technology Facilities Council grant
ST/N000668/1, and the UK Space Agency grant ST/N00180X/1.
Funding for SDSS-III and SDSS-IV has been provided by the Alfred
P. Sloan Foundation and Participating Institutions. Additional funding
for SDSS-III comes from the National Science Foundation and the
U.S. Department of Energy Office of Science. Further information about
both projects is available at www.sdss. org. SDSS is managed by the
Astrophysical Research Consortium for the Participating Institutions
in both collaborations. In SDSS- III these include the University of
Arizona, the Brazilian Participation Group, Brookhaven National
Laboratory, Carnegie Mellon University, University of Florida, the
French Participation Group, the German Participation Group, Harvard
University, the Instituto de Astrofisica de Canarias, the Michigan
State / Notre Dame / JINA Participation Group, Johns Hopkins
University, Lawrence Berkeley National Laboratory, Max Planck
Institute for Astrophysics, Max Planck Institute for Extraterrestrial
Physics, New Mexico State University, New York University, Ohio State
University, Pennsylvania State University, University of Portsmouth,
Princeton University, the Spanish Participation Group, University of
Tokyo, University of Utah, Vanderbilt University, University of
Virginia, University of Washington, and Yale University.  The
Participating Institutions in SDSS-IV are Carnegie Mellon University,
Colorado University, Boulder, Harvard- Smithsonian Center for
Astrophysics Participation Group, Johns Hopkins University, Kavli
Institute for the Physics and Mathematics of the Universe
Max-Planck-Institut fuer Astrophysik (MPA Garching),
Max-Planck-Institut fuer Extraterrestrische Physik (MPE),
Max-Planck-Institut fuer Astronomie (MPIA Heidelberg), National
Astronomical Observatories of China, New Mexico State University, New
York University, The Ohio State University, Penn State University,
Shanghai Astronomical Observatory, United Kingdom Participation Group,
University of Portsmouth, Univer- sity of Utah, University of
Wisconsin, and Yale University.  This work made use of the facilities
and staff of the UK Sciama High Performance Computing cluster
supported by the ICG, SEPNet and the University of Portsmouth. This
research used resources of the National Energy Research Scientific
Computing Center, a DOE Office of Science User Facility supported by
the Office of Science of the U.S. Department of Energy under Contract
No. DE-AC02-05CH11231.

%
%
%  These Macros are taken from the AAS TeX macro package version 4.0.
%  Include this file in your LaTeX source only if you are not using
%  the AAS TeX macro package and need to resolve the macro definitions
%  in the BibTeX entries returned by the ADS abstract service.
%
%  For more information on the AASTeX macro package, please see the URL
%	http://www.aas.org/publications/aastex.html
%  For more information about ADS abstract server, please see the URL
%	http://adswww.harvard.edu/ads_abstracts.html
%

% Abbreviations for journals.  The object here is to provide authors
% with convenient shorthands for the most "popular" (often-cited)
% journals; the author can use these markup tags without being concerned
% about the exact form of the journal abbreviation, or its formatting.
% It is up to the keeper of the macros to make sure the macros expand
% to the proper text.  If macro package writers agree to all use the
% same TeX command name, authors only have to remember one thing, and
% the style file will take care of editorial preferences.  This also
% applies when a single journal decides to revamp its abbreviating
% scheme, as happened with the ApJ (Abt 1991).

\def\jnl@style{\it}
%commente par Seb
\def\aaref@jnl#1{{\jnl@style#1}}
%ref remplace par aaref pour eviter conflit...

\def\aaref@jnl#1{{\jnl@style#1}}

\def\aj{\aaref@jnl{AJ}}                   % Astronomical Journal
\def\araa{\aaref@jnl{ARA\&A}}             % Annual Review of Astron and Astrophys
\def\apj{\aaref@jnl{ApJ}}                 % Astrophysical Journal
\def\apjl{\aaref@jnl{ApJ}}                % Astrophysical Journal, Letters
\def\apjs{\aaref@jnl{ApJS}}               % Astrophysical Journal, Supplement
\def\ao{\aaref@jnl{Appl.~Opt.}}           % Applied Optics
\def\apss{\aaref@jnl{Ap\&SS}}             % Astrophysics and Space Science
\def\aap{\aaref@jnl{A\&A}}                % Astronomy and Astrophysics
\def\aapr{\aaref@jnl{A\&A~Rev.}}          % Astronomy and Astrophysics Reviews
\def\aaps{\aaref@jnl{A\&AS}}              % Astronomy and Astrophysics, Supplement
\def\azh{\aaref@jnl{AZh}}                 % Astronomicheskii Zhurnal
\def\baas{\aaref@jnl{BAAS}}               % Bulletin of the AAS
\def\jrasc{\aaref@jnl{JRASC}}             % Journal of the RAS of Canada
\def\memras{\aaref@jnl{MmRAS}}            % Memoirs of the RAS
\def\mnras{\aaref@jnl{MNRAS}}             % Monthly Notices of the RAS
\def\pra{\aaref@jnl{Phys.~Rev.~A}}        % Physical Review A: General Physics
\def\prb{\aaref@jnl{Phys.~Rev.~B}}        % Physical Review B: Solid State
\def\prc{\aaref@jnl{Phys.~Rev.~C}}        % Physical Review C
\def\prd{\aaref@jnl{Phys.~Rev.~D}}        % Physical Review D
\def\pre{\aaref@jnl{Phys.~Rev.~E}}        % Physical Review E
\def\prl{\aaref@jnl{Phys.~Rev.~Lett.}}    % Physical Review Letters
\def\pasp{\aaref@jnl{PASP}}               % Publications of the ASP
\def\pasj{\aaref@jnl{PASJ}}               % Publications of the ASJ
\def\qjras{\aaref@jnl{QJRAS}}             % Quarterly Journal of the RAS
\def\skytel{\aaref@jnl{S\&T}}             % Sky and Telescope
\def\solphys{\aaref@jnl{Sol.~Phys.}}      % Solar Physics
\def\sovast{\aaref@jnl{Soviet~Ast.}}      % Soviet Astronomy
\def\ssr{\aaref@jnl{Space~Sci.~Rev.}}     % Space Science Reviews
\def\zap{\aaref@jnl{ZAp}}                 % Zeitschrift fuer Astrophysik
\def\nat{\aaref@jnl{Nature}}              % Nature
\def\iaucirc{\aaref@jnl{IAU~Circ.}}       % IAU Cirulars
\def\aplett{\aaref@jnl{Astrophys.~Lett.}} % Astrophysics Letters
\def\apspr{\aaref@jnl{Astrophys.~Space~Phys.~Res.}}
                % Astrophysics Space Physics Research
\def\bain{\aaref@jnl{Bull.~Astron.~Inst.~Netherlands}} 
                % Bulletin Astronomical Institute of the Netherlands
\def\fcp{\aaref@jnl{Fund.~Cosmic~Phys.}}  % Fundamental Cosmic Physics
\def\gca{\aaref@jnl{Geochim.~Cosmochim.~Acta}}   % Geochimica Cosmochimica Acta
\def\grl{\aaref@jnl{Geophys.~Res.~Lett.}} % Geophysics Research Letters
\def\jcp{\aaref@jnl{J.~Chem.~Phys.}}      % Journal of Chemical Physics
\def\jgr{\aaref@jnl{J.~Geophys.~Res.}}    % Journal of Geophysics Research
\def\jqsrt{\aaref@jnl{J.~Quant.~Spec.~Radiat.~Transf.}}
                % Journal of Quantitiative Spectroscopy and Radiative Transfer
\def\memsai{\aaref@jnl{Mem.~Soc.~Astron.~Italiana}}
                % Mem. Societa Astronomica Italiana
\def\nphysa{\aaref@jnl{Nucl.~Phys.~A}}   % Nuclear Physics A
\def\physrep{\aaref@jnl{Phys.~Rep.}}   % Physics Reports
\def\physscr{\aaref@jnl{Phys.~Scr}}   % Physica Scripta
\def\planss{\aaref@jnl{Planet.~Space~Sci.}}   % Planetary Space Science
\def\procspie{\aaref@jnl{Proc.~SPIE}}   % Proceedings of the SPIE
\def\jcap{\aaref@jnl{J. Cosmology Astropart. Phys.}}
                % Journal of Cosmology and Astroparticle Physics

\let\astap=\aap
\let\apjlett=\apjl
\let\apjsupp=\apjs
\let\applopt=\ao

\newcommand{\mpc}{\, {\rm Mpc}}
\newcommand{\kpc}{\, {\rm kpc}}
\newcommand{\hmpc}{\, h^{-1} \mpc}
\newcommand{\ihmpc}{\, h\, {\rm Mpc}^{-1}}
\newcommand{\ikms}{\, {\rm s\, km}^{-1}}
\newcommand{\kms}{\, {\rm km\, s}^{-1}}
\newcommand{\hkpc}{\, h^{-1} \kpc}
\newcommand{\lya}{Ly$\alpha$\ }
\newcommand{\lyb}{Lyman-$\beta$\ }
\newcommand{\lyaf}{Ly$\alpha$ forest}
\newcommand{\lr}{\lambda_{{\rm rest}}}
\newcommand{\bF}{\bar{F}}
\newcommand{\bS}{\bar{S}}
\newcommand{\bC}{\bar{C}}
\newcommand{\bB}{\bar{B}}
\newcommand{\vdF}{{\mathbf \delta_F}}
\newcommand{\vdS}{{\mathbf \delta_S}}
\newcommand{\vdf}{{\mathbf \delta_f}}
\newcommand{\vdn}{{\mathbf \delta_n}}
\newcommand{\vdC}{{\mathbf \delta_C}}
\newcommand{\vdX}{{\mathbf \delta_X}}
\newcommand{\xrei}{x_{rei}}
\newcommand{\lrmin}{\lambda_{{\rm rest, min}}}
\newcommand{\lrmax}{\lambda_{{\rm rest, max}}}
\newcommand{\lmin}{\lambda_{{\rm min}}}
\newcommand{\lmax}{\lambda_{{\rm max}}}
\newcommand{\hi}{\mbox{H\,{\scriptsize I}\ }}
\newcommand{\heii}{\mbox{He\,{\scriptsize II}\ }}
\newcommand{\vp}{\mathbf{p}}
\newcommand{\vq}{\mathbf{q}}
\newcommand{\vxperp}{\mathbf{x_\perp}}
\newcommand{\vkperp}{\mathbf{k_\perp}}
\newcommand{\vrperp}{\mathbf{r_\perp}}
\newcommand{\vx}{\mathbf{x}}
\newcommand{\vy}{\mathbf{y}}
\newcommand{\vk}{\mathbf{k}}
\newcommand{\vR}{\mathbf{r}}
\newcommand{\tdtwo}{\tilde{b}_{\delta^2}}
\newcommand{\tstwo}{\tilde{b}_{s^2}}
\newcommand{\tbthree}{\tilde{b}_3}
\newcommand{\tadtwo}{\tilde{a}_{\delta^2}}
\newcommand{\tastwo}{\tilde{a}_{s^2}}
\newcommand{\tabthree}{\tilde{a}_3}
\newcommand{\vnabla}{\mathbf{\nabla}}
\newcommand{\tpsi}{\tilde{\psi}}
\newcommand{\vv}{\mathbf{v}}
\newcommand{\fnl}{{f_{\rm NL}}}
\newcommand{\tfnl}{{\tilde{f}_{\rm NL}}}
\newcommand{\gnl}{g_{\rm NL}}
\newcommand{\orderfour}{\mathcal{O}\left(\delta_1^4\right)}
\newcommand{\SDSSPF}{\cite{2006ApJS..163...80M}}
\newcommand{\PF}{$P_F^{\rm 1D}(k_\parallel,z)$}
\newcommand\ionalt[2]{#1$\;${\scriptsize \uppercase\expandafter{\romannumeral #2}}}%  
\newcommand{\vxone}{\mathbf{x_1}}
\newcommand{\vxtwo}{\mathbf{x_2}}
\newcommand{\vRot}{\mathbf{r_{12}}}
\newcommand{\cm}{\, {\rm cm}}

\bibliography{draft}

\label{lastpage}

\newpage

\appendix

\end{document}